\documentclass[pre,twocolumn,floatfix,superscriptaddress]{revtex4}
\usepackage{graphicx,amsfonts,amssymb,amsmath, hyperref}
\usepackage[applemac]{inputenc}

\newif\ifhyper
\hypertrue
\ifhyper
\hypersetup{
  citecolor = {green},
  colorlinks = {true}, 
  urlcolor = {blue} 
}
\fi

\newlength{\ldag}
\begin{document}

\title{Temperature Expansions in the Square-Shoulder Fluid I: The Wiener-Hopf Function}

\author{O. Coquand} 
\email{oliver.coquand@dlr.de}
\affiliation{Institut f\"ur Materialphysik im Weltraum, Deutsches Zentrum f\"ur Luft- und Raumfahrt (DLR), 51170 K\"oln, Germany}

\author{M. Sperl} 
\email{matthias.sperl@dlr.de}
\affiliation{Institut f\"ur Materialphysik im Weltraum, Deutsches Zentrum f\"ur Luft- und Raumfahrt (DLR), 51170 K\"oln, Germany}
\affiliation{Institut f\"ur Theoretische Physik, Universit\"at zu K\"oln, 50937 K\"oln, Germany}


\begin{abstract}

	We investigate the spatial structure of dense square-shoulder fluids.
	To this end we derive analytical perturbative solutions of the Ornstein-Zernike equation in the low- and high-temperature limits as expansions around the known hard sphere solutions.
	We then discuss the suitability of perturbative approaches in relation to the Ornstein-Zernike equation.
	Our analytical expressions are shown to reproduce reasonably well numerical data in appropriate regimes.

\end{abstract}

\maketitle

\section{Introduction}

	The static structure factor is a function that characterizes the internal structure of a fluid.
	It is also directly accessible to experiments, notably via scattering methods.
	It carries quite a lot of information on the thermodynamic properties of the fluid \cite{Hansen06} --- it can be used to derive the equation of state --- and at least part of
	the dynamics of arrest in the supercooled regime \cite{Goetze08,Sperl10,Das13,Gnan14}.

	However, from a theoretical point of view, computing a structure factor reveals to be particularly difficult,
	and few analytical results have been established so far.
	One of the first systems where a structure factor was computed analytically is the hard-sphere system, where a first pathway has been designed by Wertheim and Thiele \cite{Wertheim63,Thiele63,Wertheim64},
	then followed by Baxter \cite{Baxter68} who proposed an alternative method where the full complexity of the structure factor is captured by a remarkably simple function.
	These works have then triggered a series of studies to refine our understanding of the hard-sphere fluid's structure
	\cite{Leutheusser84,Yuste91,Ripoll95,Yuste96,Robles04,Robles06,Tejero07,Rohrmann07,Robles07,Rohrmann08,	Adda08a,Adda08b,Adda08c,Katzav19}.

	One of the simplest generalizations of the hard-sphere potential is the square-shoulder potential, which consists of adding a finite region of constant positive potential outside of the hard-core.
	It shares most properties with the hard sphere potential: it is finite-range, fully repulsive, and prevents particles from getting too close to each other.
	Moreover, this potential has two natural hard-sphere limits: when the shoulder potential is very strong --- or equivalently the temperature is very low --- the outer-core becomes hard,
	on the other hand, when it becomes very soft --- or equivalently when the temperature is very high --- only the hard inner-core plays a significant role.
	Despite these properties, and a significant effort towards the theoretical understanding of this potential by use of various methods -- improved mean-spherical approximation \cite{Hlushak13},
	thermodynamic perturbation theories \cite{Lang99,Khanpour13,Kim14}, Rational Fraction Approximation \cite{Yuste11,Haro16}),
	no explicit analytical expression of the associated structure factor has been proposed yet, to the best of our knowledge.
	The closest result to such a solution has been gotten by use of the Rational Fraction Approximation (RFA) \cite{Yuste11,Haro16}, which in the spirit of the first works on the hard-sphere
	system \cite{Wertheim63,Thiele63,Wertheim64} is based on truncations of functions in Laplace space.
	More precisely, one function related to the Laplace transform of the pair-correlation function $g(r)$ is expressed as a Pad\'e approximant, the coefficients of which are then fixed by
	physical constraints.
	However, in the square-shoulder case, one of these constraints equations is transcendental and has to be solved numerically.

	Our goal in this study is to get an understanding of how complexity emerges in this seemingly simple system, by an investigation of the behavior of the structure factor of the
	square-shoulder fluid in the vicinity of its hard-sphere limits where its expression is known.
	Starting from Baxter's solution \cite{Baxter68}, which is the hard-sphere solution with the simplest formulation, we build the lowest order corrections to the hard-sphere behavior in a perturbative
	fashion, both in the low temperature limit where, compared to the particle's typical energy, the shoulder appears quasi-hard, and in the high-temperature regime where the soft core potential barrier is
	small compared to their typical kinetic energy.
	This allows us to highlight how the structure of the Ornstein-Zernike equation prevents the construction of a simple solution in this latter regime, already at lowest order.
	
	The paper is organized as follows: in a first section, we recall the method of Baxter.
	In the two following sections we build respectively the low- and high-temperature expansions around that solution,
	and discuss the properties of the perturbative series.
	Finally, we compare our results to various sets of numerical data.
	The thermodynamic properties described by those structure factors are described in a second companion paper \cite{Coquand19}

\section{Reminder: the hard-spheres system}

		First, we recall the Baxter's derivation of the structure factor of the hard-sphere fluid \cite{Baxter68}.
		The diameter of the hard spheres is called $R$.
		Remembering that in a fluid the structure factor $S(q)$ has no singularity, the Wiener-Hopf factorization can be used to write it as follows:
		\begin{equation}
		\label{eqS}
			S(q)=\big(Q(q)Q(-q)\big)^{-1}\ .
		\end{equation}
		Additionaly, $Q$ is a real function \cite{Baxter68}.
		In an isotropic fluid, $S$ does not depend on the direction of the wave vector $q$, and the Wiener-Hopf function $Q$ is related to its direct space counterpart by the following relation:
		\begin{equation}
		\label{eqQS}
			Q(q)=1-2\pi\rho\int_0^{+\infty}dr\,e^{iqr}Q(r)\ ,
		\end{equation}
		where $\rho=N/V$ is the fluid's density, $N$ the number of particles and $V$ the volume of the system.

		Finally, the Ornstein-Zernike equations can be rewritten in terms of the Wiener-Hopf function $Q(r)$ \cite{Baxter68}:
		\begin{equation}
		\label{eqOZ}
			\begin{split}
				& r c(r) = -Q'(r)+2\pi\rho\int_r^{+\infty}ds\,Q'(s)Q(s-r) \\
				& r h(r) = -Q'(r)+2\pi\rho\int_0^{+\infty}ds\,(r-s)h(|r-s|)Q(s)\ ,
			\end{split}
		\end{equation}
		where $c(r)$ is the fluid's direct correlation function, and $h$ is related to the pair correlation function $g$ by $h(r)=g(r)-1$.

		If the Ornstein-Zernike equation are closed by use of the Percus-Yevick equation:
		\begin{equation}
		\label{eqPYA}
			c(r)=(1-e^{U(r)/k_BT})g(r)\ ,
		\end{equation}
		where $U(r)$ is the particle pair potential, we also get $c(r)=0$ as long as $r>R$.
		Then Eq.~(\ref{eqOZ}) naturally leads to choose $Q(r)=0$ is this region as well.
		As a result, all integrals in Eq.~(\ref{eqOZ}) evaluate on a finite domain.

		Because hard particles cannot overlap, $h(r)=-1$ for $r\leqslant R$.
		We therefore need to compute $Q(r)$ only in a region where $h(r)$ is known exactly, what greatly simplifies the search for a solution.
		Two successive derivatives can indeed be applied to the second equation in Eq.~(\ref{eqOZ}) to yield:
		\begin{equation}
		\label{eqQHS}
			Q^{(3)}(r)=0\ ,
		\end{equation}
		that is, $Q$ is a polynomial of degree 2 in $r$.

		Finally, plugging this condition back into Eq.~(\ref{eqOZ}), with the additional requirement that $Q$ is continuous at $r=R$, gives the final result:
		\begin{equation}
		\label{eqQBax}
			Q(r)=\frac{a_b}{2}\,r^2+b_b\,r+c_b\ ,
		\end{equation}
		where the three coefficients can be written in terms of $R$ and the packing fraction $\varphi=\pi\rho R^3/6$ :
		\begin{equation}
		\label{eqCOBaxter}
			\begin{split}
				& a_b = \frac{1+2\varphi}{(1-\varphi)^2}\\
				& b_b = -\frac{3R\,\varphi}{2(1-\varphi)^2}\\
				& c_b = -\frac{R^2}{2(1-\varphi)}\ .
			\end{split}
		\end{equation}

		The structure factor is then easily deduced from Eq.~(\ref{eqQS}) and Eq.~(\ref{eqS}).
		The greatest strength of this formalism is that a function $S(q)$ with an \textit{a priori} very involved expression is completely expressed in terms
		of a simple polynomial of degree two with only two independent coefficients.
		Therefore, such a method appears to be a promising candidate to investigate small deviations from the hard-sphere potential.
		A similar study has been conducted previously on the square-well potential \cite{Dawson00}, which is similar to the square-shoulder potential from this perspective.

\section{Low-temperature expansion}

		The square-shoulder potential is defined by addition of a region of constant, positive potential $U_0$ to the hard-sphere case:
		\begin{equation}
		\label{eqSqSh}
			U(r)=\left\{\begin{split}
				& +\infty \ , \\
				& U_0     \ , \\
				& 0       \ ,
			\end{split}\right. \!\!\!\!\!\!\!\!\!\!\!\!\!\!\!\!\!\!\!\!\!\!\!\!\!\!\!\!\!\!\!\!\!\!\!\!\!\!\!\!\!\!\!\!\!\!\!\!\!\!\!\!
			\begin{split}
				& 0\leqslant r<R \\
				& R \leqslant r < d \\
				& d\leqslant r \ ,
			\end{split}
		\end{equation}
		where $d=\lambda R$ is the outer-core diameter.
		The packing fraction of the outer core $\phi=\pi\rho d^3/6$ can also be defined.
		The square-shoulder potential is one of the simplest generalizations of the hard-sphere potential with one additional characteristic length scale.
		In particular, it shares the properties of being entirely repulsive, and short-range.

		In order to be able to solve the Ornstein-Zernike equation Eq.~(\ref{eqOZ}) with the Percus-Yevick closure, Eq.~(\ref{eqPYA}), we need simplifications in the three regions
		of Eq.~(\ref{eqSqSh}).
		By analogy with the hard-sphere case, we know that inside the hard-core $g(r)=0$, and in the outside region where $U(r)=0$, the direct correlation function $c(r)$ is also equal to zero.
		We will thus assume $Q(r)=0$ for $r>d$ (it is a trivial solution of Eq.~(\ref{eqOZ})).
		Inside the soft core however, additional assumptions are needed.

		In the low-temperature limit $U_0\gg k_BT$, the contact value $g(R^+)$ is higher than in the corresponding hard-spheres system.
		Indeed, the hard-sphere contact value in the Percus-Yevick approximation is:
		\begin{equation}
		\label{eqCV}
			g(R^+)=\frac{2+\varphi}{2(1-\varphi)^2}\ ,
		\end{equation}
		which gets bigger and bigger as $\varphi$ approaches 1.
		As the temperature is decreased, the potential in Eq.~(\ref{eqSqSh}) resembles the one for hard-spheres of diameter $d$, whose packing fraction is not $\varphi$,
		but $\phi=\lambda^3\varphi>\varphi$.
		Thus this contact value can grow quite big, but is always finite if we do not allow $\phi$ to grow bigger than 1.

		Then, for larger values of $r$ inside the shoulder,
		after a very sharp decrease, $g(r)$ saturates to a value which is all the more small that the temperature is small.
		This can be explained in the following way: if $U_0\gg k_BT$, very few particles have the possibility to interpenetrate in the shoulder region.
		The expected form of $g(r)$ is then that of a low density gas, which saturates to the value of the average fraction $p$ of particles that are able
		to cross the potential barrier.

		Since the decrease of $g(r)$ is very sharp, the contact value contributes little to the integrals in the Ornstein-Zernike equations.
		For our purpose, we will thus approximate $g$ in the following way:
		\begin{equation}
		\label{eqLT}
			g(r)=p\ ,\quad R<r<d \ .
		\end{equation}
		There can be various ways to define the constant $p$, but its precise form has little impact on the quantitative results.
		For the moment, we will therefore leave it unspecified.
		Since $p$ is all the more small as $T$ is small, we will use it as a small parameter for our expansion
		\footnote{Indeed, in most cases, the form of $p$ will depend on $T$ through exponential terms that cannot be Taylor expanded in this regime.}.

		Plugging Eq.~(\ref{eqLT}) back into the Ornstein-Zernike equation Eq.~(\ref{eqOZ}), it is convenient to split the Wiener-Hopf function
		into four parts:
		\begin{equation}
		\label{eqQsplit}
			Q(r)=\left\{\begin{split}
				& Q_I(r)  \,\,\,\,  \ , \quad 0\leqslant r <d-R \\
				& Q_{II}(r)\,\,     \ , \quad d-R\leqslant r < R \\
				& Q_{III}(r)        \ , \quad R\leqslant r < d \\
				& 0 \quad\quad\quad\!\ , \quad r\geqslant d \ . 
			\end{split}\right.
		\end{equation}
		The Eq.~(\ref{eqOZ}) then rewrites:
		\begin{equation}
		\label{eqLT1}
			\begin{split}
				& (p-1)r = -Q_{III}'(r)+2\pi\rho\int_0^d Q(s)(s-r)ds           \\
				& \qquad\qquad\qquad +2\pi\rho\,p\int_0^{r-R}Q_I(s)(r-s)ds     \\
				&     -r = -Q_{II} '(r)+2\pi\rho\int_0^d Q(s)(s-r)ds           \\
				&     -r = -Q_I    '(r)+2\pi\rho\int_0^d Q(s)(s-r)ds           \\
				& \qquad\qquad\qquad +2\pi\rho\,p\int_{r+R}^dQ_{III}(s)(r-s)ds \ .
			\end{split}
		\end{equation}
		Note that due to the additional integral terms in the first and last equations, the condition Eq.~(\ref{eqQHS}) only holds in the region $II$ (for $r\in[d-R;R]$).
		Acting with two derivatives on these equations leads to the following set of coupled differential equations:
		\begin{equation}
		\label{eqLT2}
			\begin{split}
				& Q_{III}^{'''}(r) = 2\pi\rho\,p\big(Q_I(r+R)+R\,Q_I'(r+R)\big) \\
				& Q_{I}^{'''}(r) = -2\pi\rho\,p\big(Q_{III}(r-R)+R\,Q_{III}'(r-R)\big)\ .
			\end{split}
		\end{equation}
		As the left-hand side of these equations is proportional to the small parameter $p$, we can already anticipate that $Q_I$ and $Q_{III}$ can be written as a polynomial of degree two
		plus a small correction.
		The equations Eq.~(\ref{eqLT2}) can be solved exactly (the details are given in Appendix \ref{ALTex}).

		The main results are as follows: both functions can be expressed as a function of six roots $\big\{X_i\big\}_{i\in[\![1;6]\!]}$,
		and generic coefficients $\big\{Y_i^I\big\}_{i\in[\![1;6]\!]}$ and $\big\{Y_i^{III}\big\}_{i\in[\![1;6]\!]}$ to be determined by the boundary conditions, in the following way:
		\begin{equation}
		\label{eqQLTY}
			Q_{III}(r)=\sum_{i=1}^6 Y_i^{III}\,e^{X_ir}, \ Q_{I}(r)=\sum_{i=1}^6 Y_i^{I}\,e^{X_ir}\ .
		\end{equation}
		The Eq.~(\ref{eqLT2}) imposes the relation:
		\begin{equation}
			\left\{\begin{split}
				&+\text{i}Y^I_j-Y^{III}_j=0 \ , \quad j\leqslant3\\
				&-\text{i}Y^I_j-Y^{III}_j=0 \ , \quad j\geqslant4\ ,
			\end{split}\right.
		\end{equation}
		so that we can restrict ourselves to $\big\{Y_i^{III}\big\}_{i\in[\![1;6]\!]}$ , that we will simply denote
		$\big\{Y_i\big\}_{i\in[\![1;6]\!]}$ .
		Moreover, the Wiener-Hopf function $Q$ must be real, therefore for $i\leqslant3$,
		\begin{equation}
			Y_i=Y_{i+3}^*\ ,
		\end{equation}
		which ensures that the number of constraints from the boundary conditions is sufficient to solve completely Eq.~(\ref{eqLT2}).

		The roots $X_i$ have the following behavior when $p$ is small:
		\begin{equation}
			X_i\underset{p\rightarrow0}{\sim}O(p^{1/3})\ ,
		\end{equation}
		so that the general solution Eq.~(\ref{eqQLTY}) expanded at order $O(p)$ is a polynomial of degree 3 in $r$.

		Finally, the Ornstein-Zernike equation Eq.~(\ref{eqLT1}) can be solved using the following ansatz:
		\begin{equation}
		\label{eqQLT}
			Q_i(r)=e_i\frac{r^3}{6}+a_i\frac{r^2}{2}+b_i\,r+c_i\ ,
		\end{equation}
		where $i\in\{I,II,III\}$, and $e_{II}=0$.
		For the sake of simplicity, let us decompose each coefficient according to its $p$ expansion:
		$e_i=e_i^{(1)}p+O(p^2)$, $a_i=a^{(0)}+a_i^{(1)}p+O(p^2)$, $b_i=b^{(0)}+b_i^{(1)}p+O(p^2)$,
		$c_i=c^{(0)}+c_i^{(1)}p+O(p^2)$.
		For each of these coefficients, the leading order term is independent of the considered region
		and is consistent with Baxter's solution for hard-spheres of diameter $d=\lambda\, R$ :
		\begin{equation}
		\label{eqLTQ0}
			\begin{split}
				& a^{(0)}=\frac{1+2\phi}{(1-\phi)^2} \\
				& b^{(0)}= -\frac{3d \phi}{2(1-\phi)^2}\\
				& c^{(0)}= -\frac{d^2}{2(1-\phi)}\ .
			\end{split}
		\end{equation}
		The other coefficients are as follows:
		\begin{equation}
		\label{eqLTQ1}
			\begin{split}
				& e_I^{(1)}=\frac{6\phi}{d(1-\phi)} \\
				& a_I^{(1)}=\frac{\phi}{\lambda^4(1-\phi)^2}
				\left[\lambda^4(\phi-10)-27\phi+4\lambda(1+8\phi)\right] \\
				& b_I^{(1)}=\frac{3R(\lambda-1)\phi}{2\lambda^3(1-\phi)^2}
				\left[1+\lambda+3\lambda^2+3\lambda^3+2(\lambda-5)\phi\right] \\
				& c_I^{(1)}= \frac{R^2(\lambda-1)}{2\lambda^2(1-\phi)}
				\left[-3\phi+\lambda(\lambda+\lambda^2+\phi)\right] \ ,
			\end{split}
		\end{equation}
		\begin{equation}
		\label{eqLTQ2}
			\begin{split}
				& a_{II}^{(1)}=-\frac{\phi}{\lambda^4(1-\phi)^2}
				\left[27\phi+\lambda^4(4+5\phi)-4\lambda(1+8\phi)\right] \\
				& b_{II}^{(1)}=\frac{3R\phi}{2\lambda^3(1-\phi)^2}
				\left[-1+\lambda^4+2(5-6\lambda+\lambda^4)\phi\right] \\
				& c_{II}^{(1)}=\frac{R^2(\lambda^2-1)}{2\lambda^2(1-\phi)}
				\left[-3\phi+\lambda^2(1+2\phi)\right] \ ,
			\end{split}
		\end{equation}
		\begin{equation}
		\label{eqLTQ3}
			\begin{split}
				& e_{III}^{(1)}=-\frac{6\phi}{d(1-\phi)} \\
				& a_{III}^{(1)}=-\frac{\phi\left[27\phi^2-4\lambda\phi(1+8\phi)+\lambda^4(1+2\phi+6\phi^2)\right]
				}{\lambda^4(1-\phi)^2} \\
				& b_{III}^{(1)}=\frac{3R \phi\left[-1+2\lambda^2+\lambda^4+2\phi(5+\lambda(\lambda^3-\lambda-6))
				\right]}{2\lambda^3(1-\phi)^2} \\
				& c_{III}^{(1)}=\frac{R^2}{2\lambda^2(1-\phi)}
				\left[\lambda^4+\phi\big(3+2\lambda(\lambda-2)(\lambda+1)^2\big)\right] \ .
			\end{split}
		\end{equation}
		These terms are the first correction to the Wiener-Hopf function of hard-spheres once a fraction of their
		hard-core becomes soft.
		Since the solution of the Ornstein-Zernike equation is known exactly, it is to be noted that such a computation
		is easily generalized to higher orders in $p$, in which case the limiting assumption should be the form
		of $g(r)$ which is taken to be constant inside the integrals.
		It then appears that going to higher and higher orders in powers of $p$ amounts to take as an ansatz for $Q(r)$
		a piecewise polynomial of increasing degree.

\section{High-temperature expansion}

	\subsection{The Full Percus-Yevick Solution}

			We now turn to the high-temperature regime of the square-shoulder potential, Eq.~(\ref{eqSqSh}): $U_0\ll k_B T$.
			Thanks to the Percus-Yevick closure, Eq.~(\ref{eqPYA}), we already have sufficient knowledge inside the hard-core, and outside the potential range:
			as in the low-temperature case, we only need further an approximation inside the soft core $r\in[R;d]$.
			Since we are still dealing with a potential with only one additional length scale, it is convenient to use the splitting,
			Eq.~(\ref{eqQsplit}), so that the Ornstein-Zernike equation now reads:
			\begin{equation}
			\label{eqHT1}
				\begin{split}
					& rh(r) = -Q_{III}'(r)+2\pi\rho\int_0^dQ(s)(s-r)ds \\
					&\qquad\quad+2\pi\rho\int_0^{r-R}Q_I(s)(r-s)g(|r-s|)ds \\
					& -r = -Q_{II}'(r)+2\pi\rho\int_0^d Q(s)(s-r)ds \\
					& -r = -Q_I'(r)+2\pi\rho\int_0^d Q(s)(s-r)ds \\
					&\qquad\quad+2\pi\rho\int_{r+R}^d Q_{III}(s)(r-s)g(|r-s|)ds\ .
				\end{split}
			\end{equation}

			In the region $II$, the ansatz (\ref{eqQBax}) still holds;
			for the two other regions however, we need knowledge about the pair correlation function inside the
			shoulder.
			However, contrary to the low-temperature case, where the pair correlation function approaches a small
			constant, in the high-temperature regime, $g(r)$ approaches its hard sphere value outside the hard core
			--- further corrections are of next order in an expansion in powers of $\Gamma=U_0/k_BT$ ---
			which full exact analytical expression is not known, to the best of our knowledge.
			The problem of finding an expression of the hard-sphere pair-correlation function $g(r)$ 
			within Percus-Yevick's approximation has nonetheless been solved
			explicitly by Wertheim in the region $r\in[R;2R]$ \cite{Wertheim63}, and then used in
			a number of subsequent studies \cite{Wertheim64,Yuste91,Evans93,Largo00,Trokhymchuk05,Haro06,Pieprzyk17}.

			For the sake of simplicity, we use here a different but equivalent approach: instead of giving an explicit
			expression to $g(r)$, the Percus-Yevick equation (\ref{eqPYA}), combined with Eq.~(\ref{eqOZ})
			can be used to re express $g$ as a function of $Q$ alone.
			The Eq.~(\ref{eqHT1}) becomes non-linear in $Q$, but a careful spliting of each part of the Wiener-Hopf
			function in the $\Gamma$ expansion, combined with the knowledge of the order $O(\Gamma^0)$ which is
			nothing	but Baxter's solution, Eq.~(\ref{eqQBax}), is sufficient to solve the equations.
			The details are given in Appendix \ref{AHTPYA}.

			The main result is that $Q_{III}$ can still be expressed in the form:
			\begin{equation}
				Q_{III}=\sum_{i=1}^6 Y_i e^{X_i r}\ ,
			\end{equation}
			where the $X_i$'s are the roots of some specific polynomial.
			However, their high-temperature expansion has the form:
			\begin{equation}
				X_i=X_i^{(0)}+\Gamma\,X_i^{(1)} +O(\Gamma^2)\ ,
			\end{equation}
			so that $Q$ does not take a polynomial form anymore:
			\begin{equation}
				Q_{III}=\sum_{i=1}^6Y_ie^{X_i^0 r}\big[1+(X_i^1r)\Gamma\big]+O(\Gamma^2)\ .
			\end{equation}
			Given the involved expressions of $X_i^{(0)}$'s, which are the roots of a polynomial of degree three,
			such an expression does not yield very useful analytical results when the boundary condition equations
			are solved.
			In the following, we design further approximation schemes, so as to get a better analytical framework to
			deal with the high-temperature regime.

			This form of $Q$'s expansion should not come as a surprise in as much as the Ornstein-Zernike equations on the Wiener-Hopf
			function $Q$, Eq.~(\ref{eqHT1}), depend explicitly on the pair correlation function in a regime where
			its analytical expression is already quite heavy \cite{Wertheim63}.
			The fact that the $\Gamma$ expansion of $Q$ is not a polynomial, even at the lowest order is a hint that
			Baxter's solution does not yield a convenient starting point for an expansion at high temperature.

	\subsection{Approximations of the pair correlation function}

			In order to get a simpler approximation of $Q(r)$, the first ingredient we need is an analytical
			approximant of $g(r)$ inside the shoulder.
			As already stated, to lowest order in $\Gamma$, we can restrict ourselves to an approximate expression
			of the hard-sphere $g(r)$ in the region near the hard-core in the Percus-Yevick approximation.

			The Ornstein-Zernike equation Eq.~(\ref{eqOZ}) relates directly the pair correlation function to the Wiener-Hopf function.
			In the hard-sphere case, it is thus possible to use Baxter's solution Eq.~(\ref{eqQBax}) to get a polynomial approximant of
			$g(r)$ in the vicinity of the hard core.
			As a matter of fact, defining $F(r)=rh(r)$, Eq.~(\ref{eqOZ}) can be written:
			\begin{equation}
			\label{eqFTayl}
				F(R+\delta r)=0+2\pi\rho\int_0^R\, Q_b(s)F(R+\delta r -s)ds \ .
			\end{equation}
			Calling $F_n(r)$ the polynomial obtained by truncating the Taylor expansion of $F$ at order $O(\delta r^n)$, we get
			an expression whose precision is typically of order $O(\delta^n)$, where $\delta=\lambda-1$.

\begin{figure}[h]
				\begin{center}
					\includegraphics[scale=0.55]{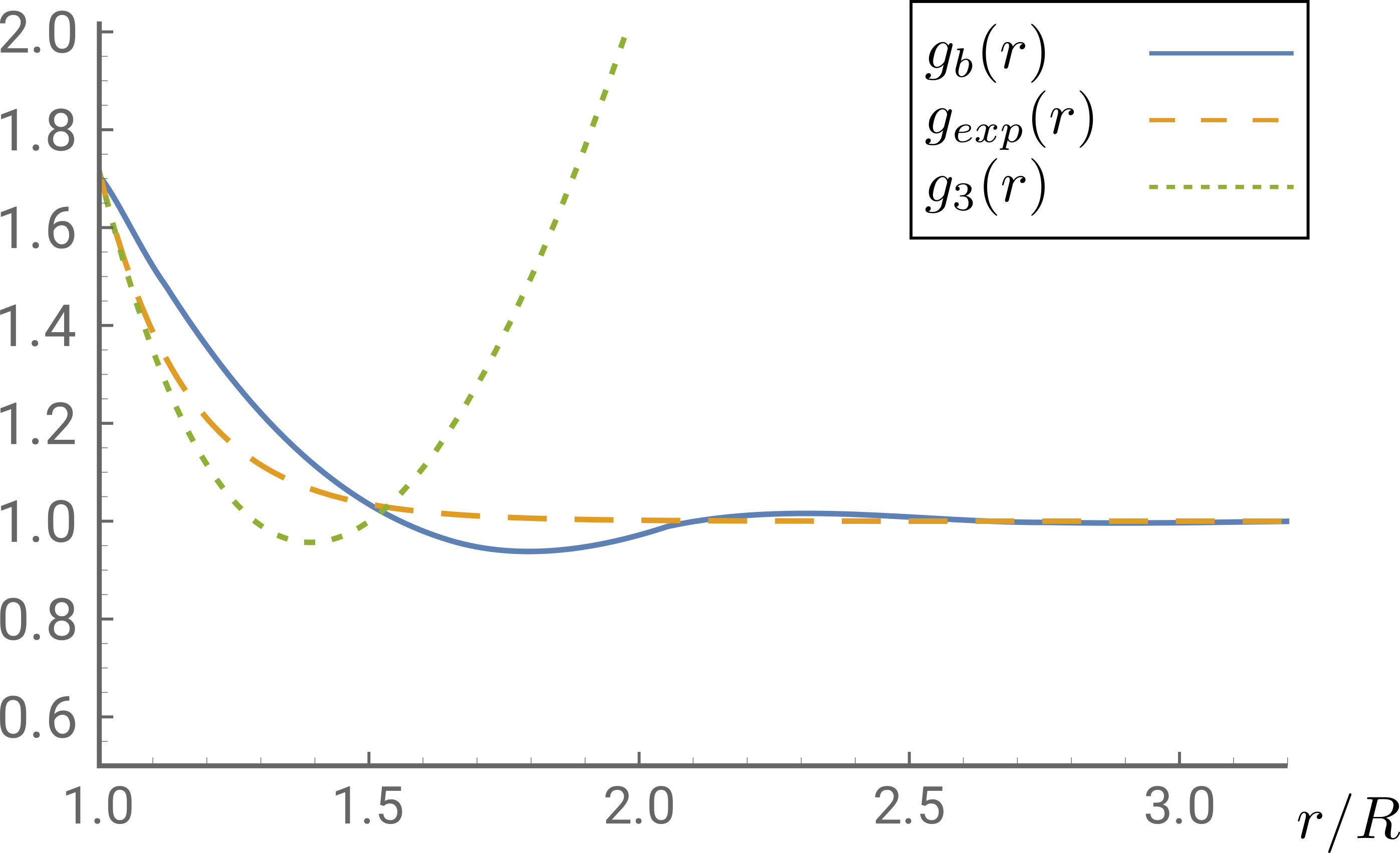}
				\end{center}
				\caption{Comparison between the Baxter's pair correlation function $g_b$ -- full line -- and different truncations, for a hard-sphere system
				of packing fraction $\varphi=0.2$.
				The dotted line corresponds to a truncation of Eq.~(\ref{eqFTayl}) at order $O(\delta^3)$, and leads to very unphysical behavior.
				The dashed line corresponds to the exponential ansatz Eq.~(\ref{eqfgexp}) used in the following.}
			\label{FigFigr}
			\end{figure}

			Such polynomial approximants are however ill-behaved and lead to unstable solutions except for very low values of $\delta$.
			Indeed, even if they reproduce well the behavior of $F$ near the core, high order terms lead very rapidly to
			excessive, positive or negative, values of $F$ (and thus $g$), at odds with the expected physical tendency, see Fig.~\ref{FigFigr}.

			To overcome this difficulty, we chose to use the simplest exponential ansatz compatible with the two first orders of the $\delta$ expansion, namely,
			\begin{equation}
			\label{eqfgexp}
				F_{exp}(r)=F_0\,\exp\left(\frac{F_1}{F_0}(r-R)\right)\ ,
			\end{equation}
			with
			\begin{equation}
				F_0=-\frac{R\varphi(-5+2\varphi)}{2(1-\varphi)^2}\ ,\quad F_1=-\frac{\varphi(10-2\varphi+\varphi^2)}{(1-\varphi)^3}\ .
			\end{equation}

			The corresponding pair correlation function $g_{exp}$ is displayed in Fig.~\ref{FigFigr}.
			It approximates well Baxter's solution in the immediate vicinity of $r=R$, and far away form it.
			However, it misses the typical oscillatory behavior.
			Clearly, the quality of such an ansatz is diminished by going to higher values of $\delta$, especially
			if $\varphi$ is also quite high, but its precision cannot be captured by a simple power of
			$\delta$, it is not anymore a small-shell expansion.

			Obviously, better approximations of $g$ can be designed -- recall that its exact expression within the
			Percus-Yevick approximation is known for $\delta\leqslant2$ \cite{Wertheim63} -- but we
			want here only to work out the simplest, yet physical, solution for the square-shoulder structure
			factor in a high-temperature expansion around Baxter's solution.
			As we show in the following, even this simplified solution is quite involved.

	\subsection{The truncated Percus-Yevick solution}

			As a consequence of the truncations of $g(r)$, we cannot work with the first equation of (\ref{eqHT1}),
			since the terms of order $O(\Gamma^0)$ do not cancel anymore -- and $Q_{III}$ is of order $O(\Gamma)$.
			Instead, the Percus-Yevick equation should be used to determine $Q_{III}$.
			In the appropriate region, it reads:
			\begin{equation}
			\label{eqPYAHT}
				\begin{split}
					&(1-e^\Gamma)(F(r)+r) =rc(r) \\
					& \ =-Q_{III}'(r)+2\pi\rho\int_r^dQ_{III}'(s)Q_{b}(s-r)ds+O(\Gamma^2)\ ,
				\end{split}
			\end{equation}
			where we have used the fact that, since $Q_{III}$ should go to zero as $\Gamma \rightarrow0$ -- this is
			the hard-sphere high-temperature limit -- the second factor in the integral can be replaced by its
			hard-sphere value $Q_b$.

			The natural way to proceed is then to take derivatives to get a linear differential equation:
			\begin{equation}
				\Gamma F^{''}=Q_{III}^{'''} +2\pi\rho\left(c_{b} Q_{III}''
				+ b_{b} Q'_{III} + a_{b} Q_{III}\right) \ .
			\end{equation}
			Note that such an expression is possible thanks to the polynomial character of $Q_b$.
			A simple solution ansatz can be found, with the form of $F$:
			\begin{equation}
				Q(r)=q_0\big(e^{q_1(r-d)}-1\big)\ ,
			\end{equation}
			but it can be shown not to fulfill the boundary conditions of the equation (\ref{eqPYAHT})
			before derivation.
			Therefore, a solution of the homogeneous equation, of the form:
			\begin{equation}
				Q_H(r)=q_0+q_1e^{x_1r}+q_2e^{x_2r}+q_3e^{x_3r}\ ,
			\end{equation}
			where $\{x_i\}$ are the roots of:
			\begin{equation}
				X^3 +2\pi\rho\left( c_{b}X^2 + b_{b}X+ a_{b}\right)\ ,
			\end{equation}
			must be added.

			We recover the same problem as in the full Percus-Yevick case: the general solution is expressed in terms of exponential terms, not reducible into polynomials by a simple $\Gamma$ expansion.
			This impedes the expression of the coefficients in $Q$ as simple functions of the boundary conditions.
			We must therefore push our approximation even further.

			In order to do so, note that if $\delta$ is not too big, the hard-sphere term in the convolution
			integral in Eq.~ (\ref{eqPYAHT}) is not too different from its constant term.
			We can thus safely replace $Q_{b}$ by $c_{b}$.
			We then do not need to take derivatives to find a linear differential equation.
			The general solution for $Q_{III}$ has the form:
			\begin{equation}
			\label{eqQ3HT}
				Q_{III}(r)=q_0\,e^{q_1 r}+q_2\,e^{q_3 r}+b_{III} r +c_{III}\ ,
			\end{equation}
			where $q_3=6\frac{\varphi}{R(1-\varphi)}$ comes from the homogeneous solution, and
			$q_1=F_1/F_0$ comes from the left-hand side term.
			The full expressions of the remaining coefficients can be found in the Appendix \ref{AHTCo}.
			As expected, $Q_{III}$ is of order $O(\Gamma)$.

			The remaining parts of the Wiener-Hopf function are then derived by plugging back into
			the Eq.~(\ref{eqHT1}) the truncated expression of $g(r)$ and the expression Eq.~(\ref{eqQ3HT}).
			As discussed before, $Q_{II}$ still has a polynomial expression,
			\begin{equation}
			\label{eqQ2HT}
				Q_{II}(r)=\frac{a_{II}}{2} r^2 + b_{II} r +c_{II}\ ,
			\end{equation}
			but $Q_I$ as well as $Q_{III}$ is not polynomial anymore:
			\begin{equation}
			\label{eqQ1HT}
				\begin{split}
					Q_I(r)= &g_I\frac{r^4}{24}+e_I\frac{r^3}{6}+a_I \frac{r^2}{2}+b_I\, r+c_I \\
					&+ q_{10}\,e^{q_1r}+ q_{11}\,e^{-q_1r} +q_{30}\,e ^{q_3r}\ .
				\end{split}
			\end{equation}

			The quite lengthy expression of the coefficients are given in the Appendix \ref{AHTCo}.
			They illustrate the complexity of the solution, even at this level of approximation.
			As expected, both $Q_I$ and $Q_{II}$ can be written as $Q_b+O(\Gamma)$.

	\subsection{Discussion}

\begin{figure}
				\begin{center}
					\includegraphics[scale=0.6]{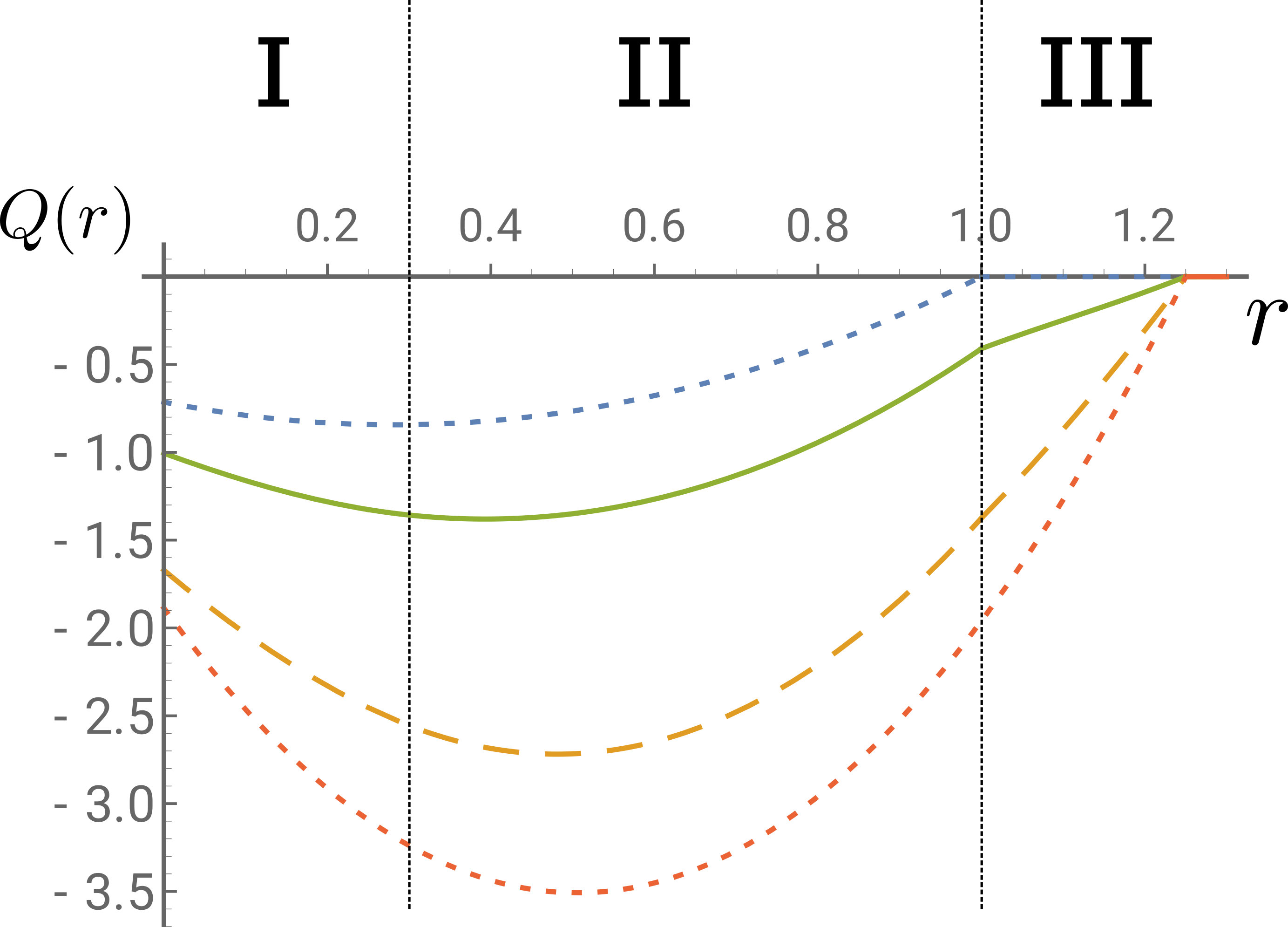}
				\end{center}
				\caption{Evolution of the Wiener-Hopf function $Q(r)$ for $\varphi=0.3$ and $R=1$.
				The dotted lines correspond to the two hard-spheres limits.
				The dashed line is the low-temperature expansion of $Q$ for $\lambda=1.25$ and $p=0.8$.
				The full line is the high-temperature expansion of $Q$ for $\lambda=1.25$ and $\Gamma=1.2$.
				The values of the temperature parameters are chosen a bit outside of the range of applicability of our formulas to emphasize
				the deformations induced by temperature.
				In particular, due to the exponential behavior of $p$ with $T$, quite large values must be taken to separate the low-temperature curve form its hard-sphere limit.}
			\label{FigQ}
			\end{figure}

			The evolution of the Wiener-Hopf function $Q$ with $r$ is represented on the Fig.~\ref{FigQ}.
			In particular, this graph displays how we get from the first hard-sphere limit to the high-temperature $Q(r)$,
			then to the low-temperature one, and finally to the second hard-sphere limit as the temperature is decreased.
			The shape of the square-shoulder $Q(r)$ appears to be quite similar to the hard-sphere one, except for a cusp located at $r=R$
			which is associated with $g(r)$ developing a second discontinuity, as is well-known from previous data (see \cite{Lang99,Yuste11} for example).

			Note that, to the contrary, the transition for $Q_I$ to $Q_{II}$ remains smooth.
			Indeed, in Eq.~(\ref{eqHT1}), as $r\rightarrow d-R$ in region $I$, the last integral term in the Ornstein-Zernike equation smoothly goes to zero.
			Thus, the values of $Q_I'$ and $Q_{II}'$ get equal in that region.

			All in all, we have investigated the possible constructions of high-temperature
			expansions around Baxter's hard-sphere solution, valid for square-shoulder
			systems with very weak shoulder potentials.
			It has been shown that the design of such an expansion turns out to be much more involved than
			in the low-temperature case.
			Even at the lowest level of approximation, the main asset of Baxter's solution, simplicity,
			is lost, alongside with the polynomial character of the Wiener-Hopf function.
			The source of these difficulties has been identified: in the high-temperature case, the knowledge
			of the form of the hard-sphere pair correlation function outside of the core plays a crucial
			role, but cannot be accurately captured by simple functions of the packing fraction
			$\varphi$.
			This will have further implications, as discussed in the companion paper.

\section{Numerical Accuracy}

	In the following section the results of our expansions are compared to various sets of numerical data.

	\subsection{The Wiener-Hopf function}

\begin{widetext}

		\begin{figure}
			\begin{center}
				\includegraphics[scale=0.6]{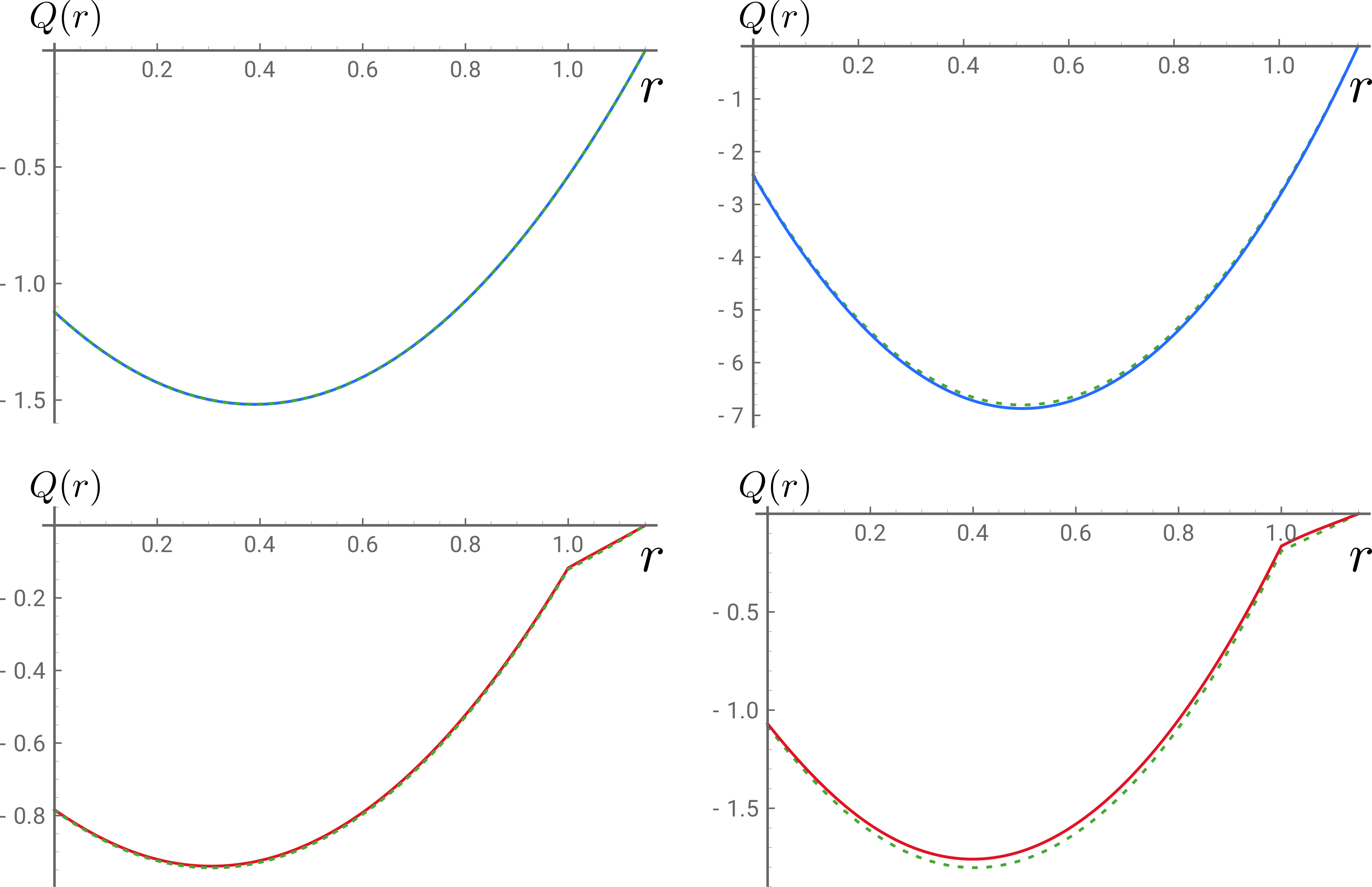}
			\end{center}
			\caption{Comparison of our expansions with the numerical solution of Eq.~(\ref{eqOZ}) used in \cite{Sperl10a} for $\lambda = 1.15$.
			The solid line is the analytical result, the dashed one corresponds to the numerical data.
			The top panels displays the low-temperature ansatz for $\Gamma=7.5$ and the bottom panels displays the high-temperature ansatz for $\Gamma=0.5$.
			In the left column $\varphi = 0.27$, in the right column $\varphi = 0.48$.}
			\label{figQtable}
		\end{figure}

		\end{widetext}

		The Wiener-Hopf function $Q$ is mostly used in a theoretical context, hence it is generally not computed
		in numerical investigations of structure problems.
		However, in \cite{Sperl10a} the author used a square-shoulder structure factor which was determined by numerically solving the $Q(r)$ version
		of the Ornstein-Zernike equation Eq.~(\ref{eqOZ}) within the Percus-Yevick approximation.
		The results are presented in Fig.~\ref{figQtable}.
		The numerical and analytical curves are almost on top of each other for each values of the parameters.
		This validates the numerical accuracy of our ansatz.

\begin{figure}
			\begin{center}
				\includegraphics[scale=0.5]{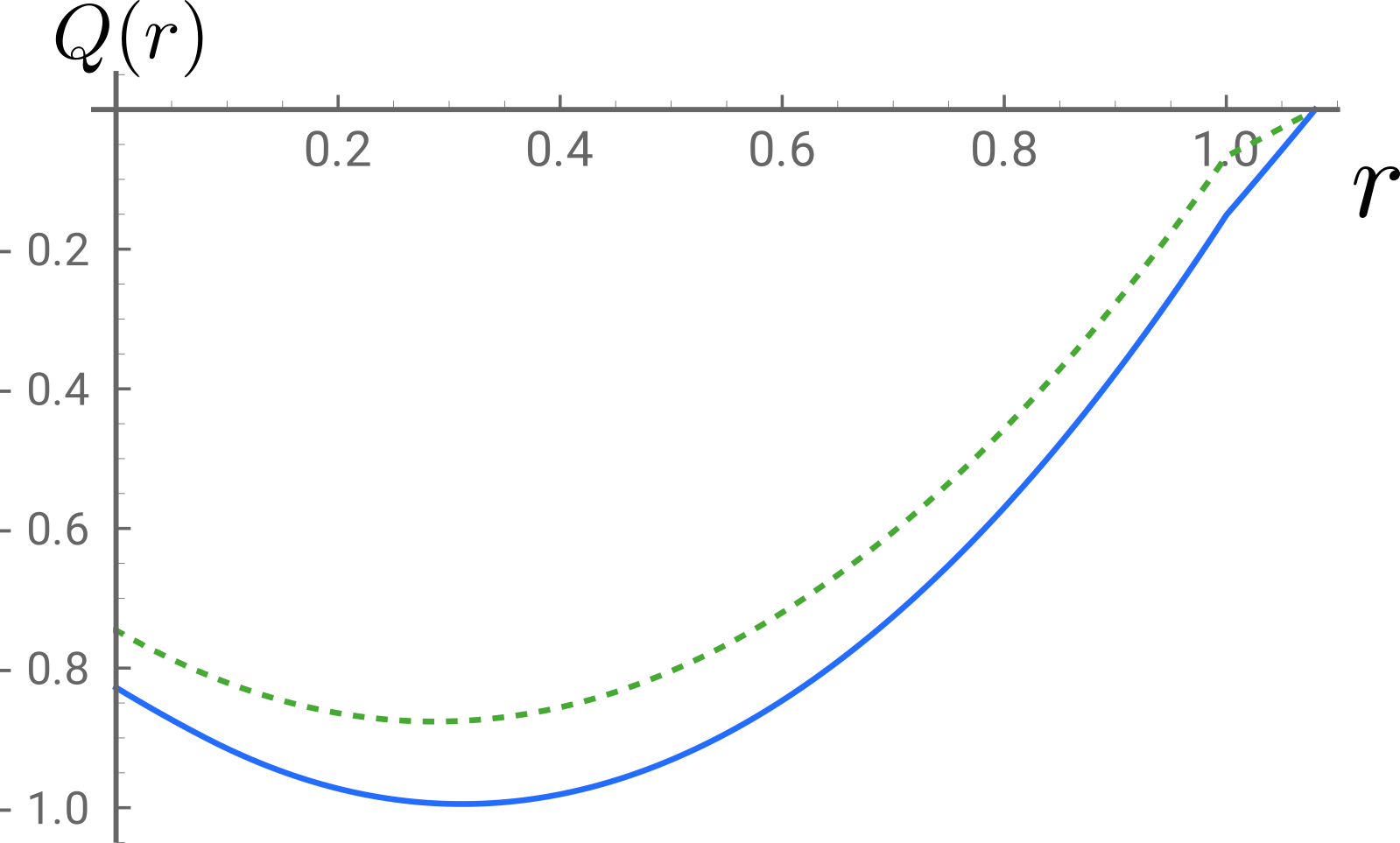}
			\end{center}
			\caption{Evolution of $Q(r)$ for $\lambda = 1.08$, $\Gamma=0.5$ and $\varphi=0.27$.
			The full line is the low-temperature result, the dashed line is the numerical solution.}
			\label{figLTnot}
		\end{figure}

		It is also instructive to look at the behaviour of the analytical ansatz out of their range of applicability.
		For example, in Fig.~\ref{figLTnot}, we represented the low-temperature ansatz for a dimensionless temperature $T=2$.
		A striking feature is that the cusp at $r=R$ is barely visible, as we already discussed in Fig.~\ref{FigQ}.
		This can be related to the fact that in this framework, we supposed that $g(r)=p$ inside the shoulder, which is not a good approximation
		anymore when the temperature rises.
		As a result the contact value $g(R^+)$, which is related to the slope difference on both sides of the cusp, is very much underestimated,
		so that the cusp is less pronounced than on the numerical result without temperature expansions.

\begin{figure}
			\begin{center}
				\includegraphics[scale=0.3]{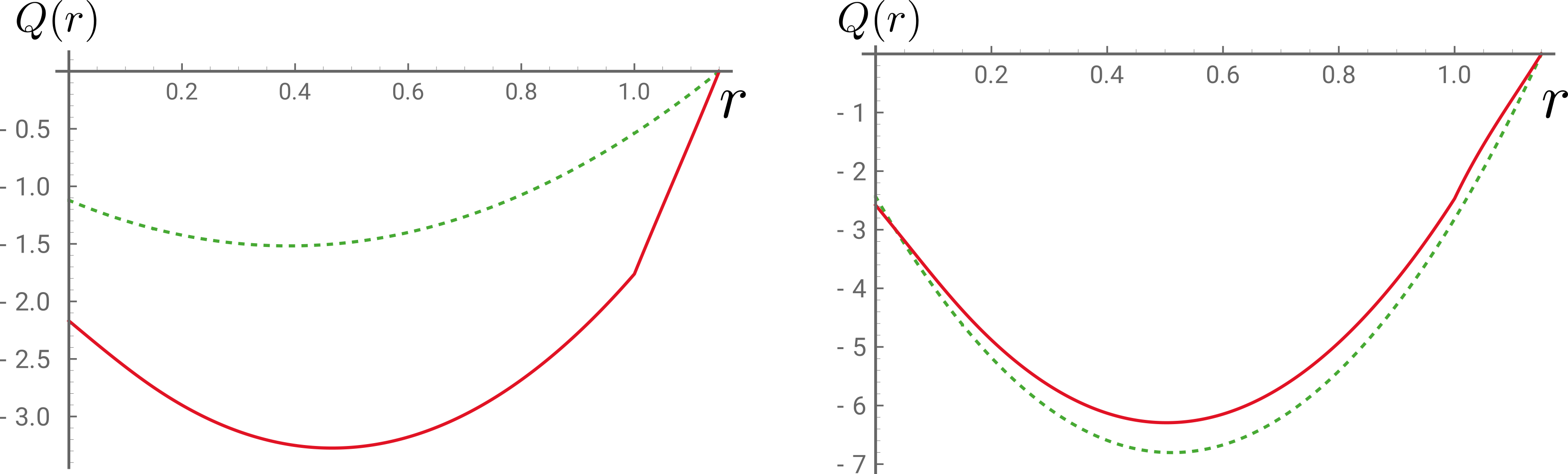}
			\end{center}
			\caption{Evolution of $Q(r)$ for $\lambda = 1.15$, $\Gamma=7.5$ and $\varphi=0.27$ (left) or $\varphi=0.48$ (right).
			The full line is the high-temperature result, the dashed line is the numerical solution.}
			\label{figHTnot}
		\end{figure}

		To the contrary, in the case of the high-temperature expansion, the cusp is overestimated at high values of $\Gamma$ (see Fig.~\ref{figHTnot}).
		As a matter of fact, the contact value $g(R^+)$ is now a linear function of $\Gamma$, which can thus become really wrong when $\Gamma$ is large
		--- see \cite{Coquand19} for the full expression of $g(R^+)$ at high temperature.
		Interestingly, as can be seen in Fig.~\ref{figHTnot}, the form of the cusp gets better when the packing fraction increases: the $\Gamma\rightarrow0$ limit of
		the contact value $g(R^+)$ is given by its hard sphere expression Eq.~(\ref{eqCV}) which increases with $\varphi$.
		Therefore, for higher packing fractions, the overestimation of the $\Gamma$ correction to the contact value at lower temperatures is
		comparatively weaker, hence a better numerical agreement.

	\subsection{Structure factor}

\begin{widetext}

		\begin{figure}
			\begin{center}
				\includegraphics[scale=0.6]{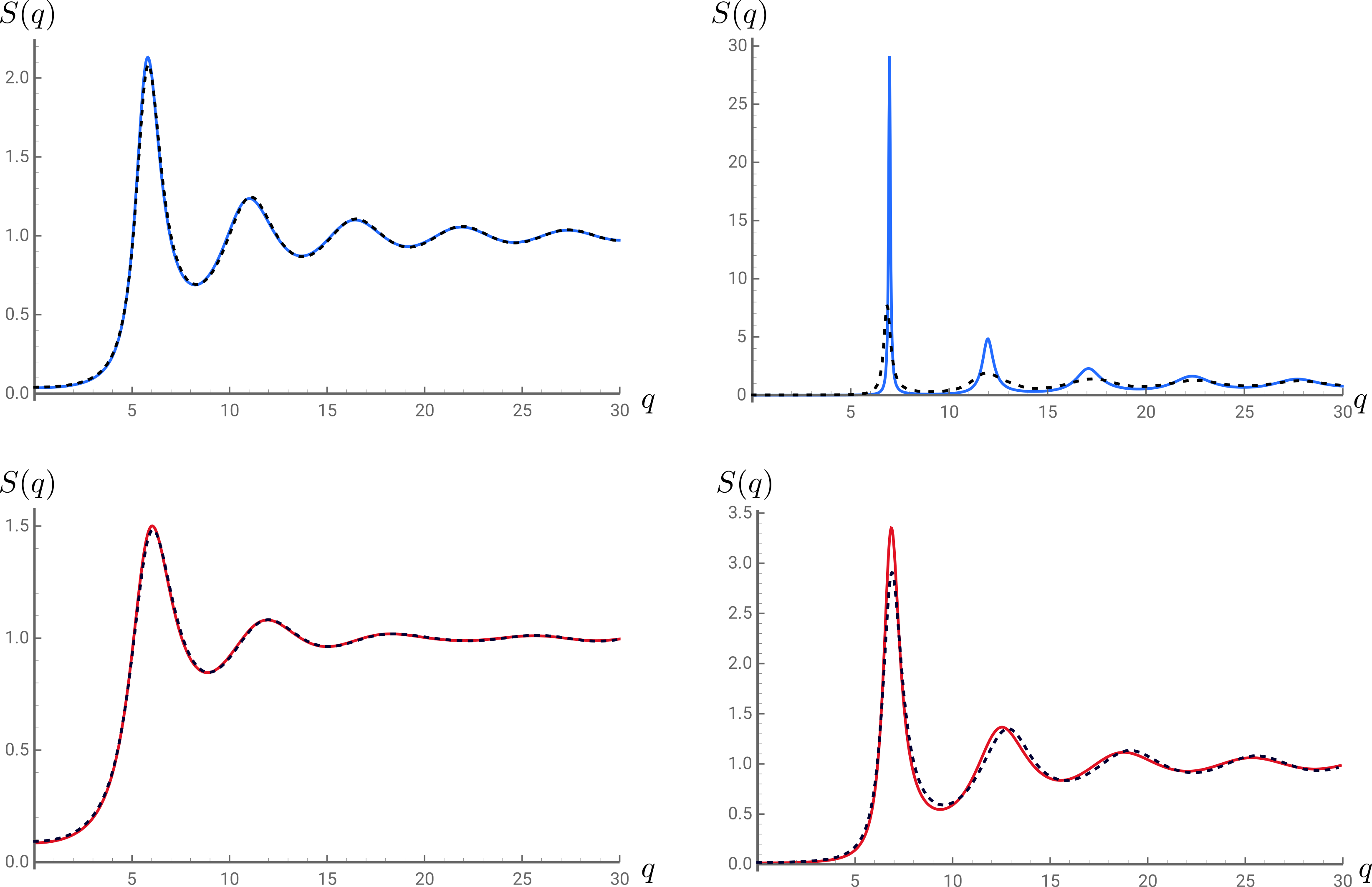}
			\end{center}
			\caption{Comparison of our expansions with the numerical solution of the Ornstein-Zernike equation with
			the Rogers-Young closure for $\lambda = 1.15$.
			The solid line is the analytical result, the dashed one corresponds to the numerical data.
			The top panels displays the low-temperature ansatz for $\Gamma=7.5$ and the bottom panels displays the high-temperature ansatz for $\Gamma=0.5$.
			In the left column $\varphi = 0.27$, in the right column $\varphi = 0.48$.}
			\label{figSqtable}
		\end{figure}

		\end{widetext}

		As a next step, we compare the structure factors built from our expansions to more realtistic data.
		Indeed, it is known that the Percus-Yevick closure, despite its analytical simplicity, leads to thermodynamical inconsistencies for example.
		A much better closure, form the point of view of numerical accuracy, is the Rogers-Young closure \cite{Rogers84},
		which is built explicitly to be thermodynamically consistent.
		It compares also very-well to numerical simulations.
		Thus, we can expect that results from the Rogers-Young closure are more accurate than ours, and use them to get an
		estimation of how good the Percus-Yevick approximation is in such system in typical ranges of parameters.
		The comparison is displayed in Fig.~\ref{figSqtable}.

		For low enough packing fractions, the agreement between the results from the two closures is good, but it deteriorates when the fluid gets denser.
		More precisely, as can be seen in the case of the high-temperature expansion, it is the first peak of the structure factor that carries the biggest error,
		which means that our structure factors tend to overestimate the enforcement of localisation of the particles in the fluid.
		Let us stress though, that on this data set $\Gamma=0.5$, which is already quite large for a $\Gamma\ll1$ epxansion.

\begin{figure}
			\begin{center}
				\includegraphics[scale=0.6]{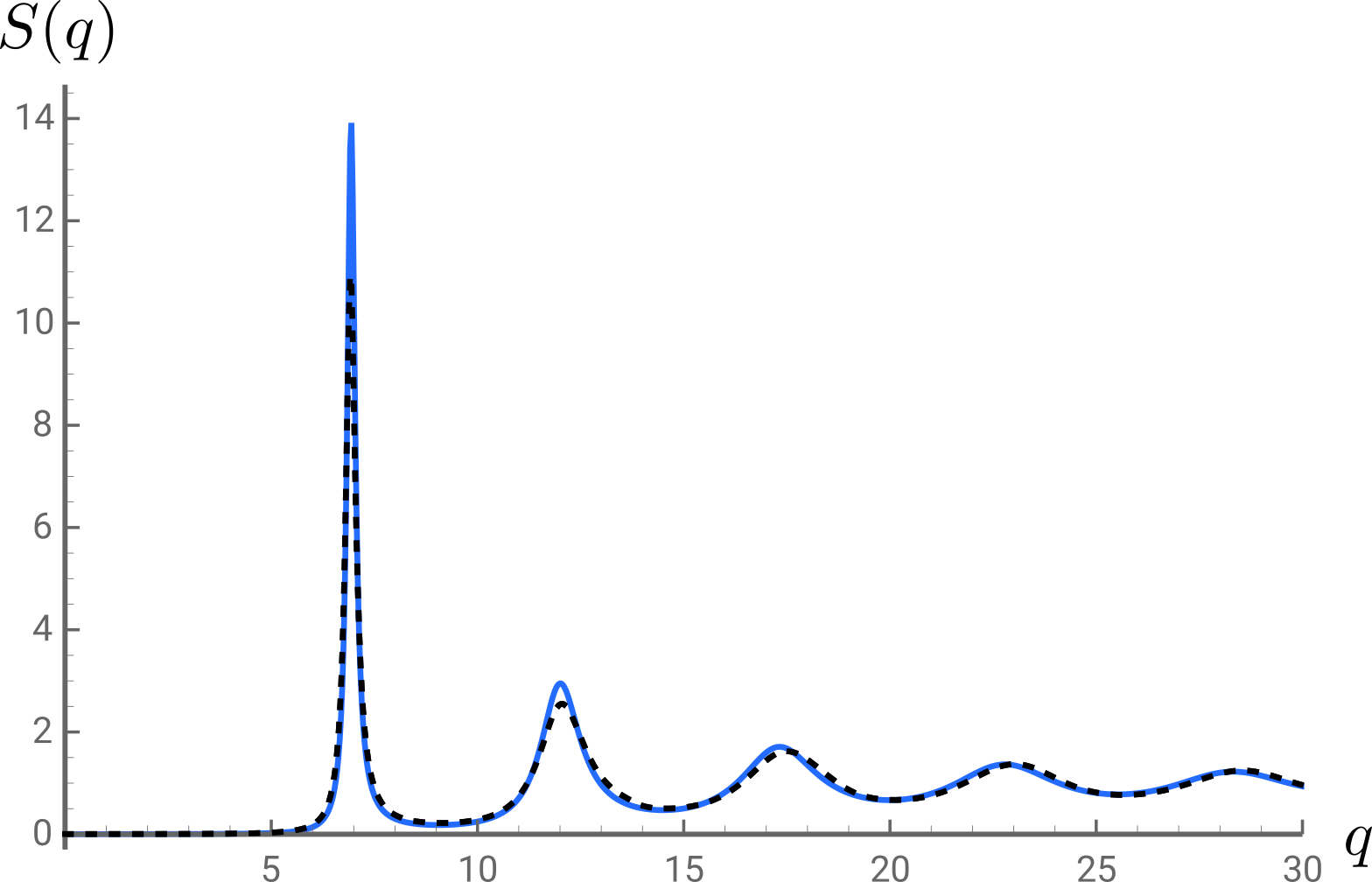}
			\end{center}
			\caption{Evolution of $S(q)$ for $\lambda = 1.12$, $\Gamma=7.5$ and $\varphi=0.48$.
			The full line is the low-temperature result, the dashed line is the numerical solution.}
			\label{figSqbetter}
		\end{figure}

		The case of the low-temperature expansion requires a closer look.
		Indeed, the agreement with the Rogers-Young data for $\varphi=0.48$ is quite poor, what could come as a surprise since the low-temperature expansion is,
		as we saw before, much better behaved than its high-temperature counterpart.
		In order to understand a bit better this result, it is necessary to recall that in the low-temperature case, the reference point for the expansion
		is a hard-sphere system with packing fraction $\phi=\varphi\,\lambda^3$.
		Therefore, not only $\varphi$, but also $\lambda$ is a crucial quantity.
		In particular, for $\varphi=0.48$ and $\lambda=1.15$, as shown on this example, the outer-core packing fraction is $\phi\simeq0.73$,
		so that the reference system can hardly be considered as being in its fluid state.
		It is to be expected that even in the hard-sphere case, the structure factor determination is not very precise at such high packing fractions.
		Extra caution is therefore needed in the low-temperature case to ensure that the reference state used in the expansion is a well-defined one.
		As a final illustration, it can be seen in Fig.~\ref{figSqbetter} that even when $\lambda$ is only lowered to $1.12$ --- in which case $\phi\simeq0.67$ ---
		the agreement between our Percus-Yevick results and the Rogers-Young structure factor is much better.

	\subsection{Pair correlation function}

\begin{widetext}

		\begin{figure}
			\begin{center}
				\includegraphics[scale=0.6]{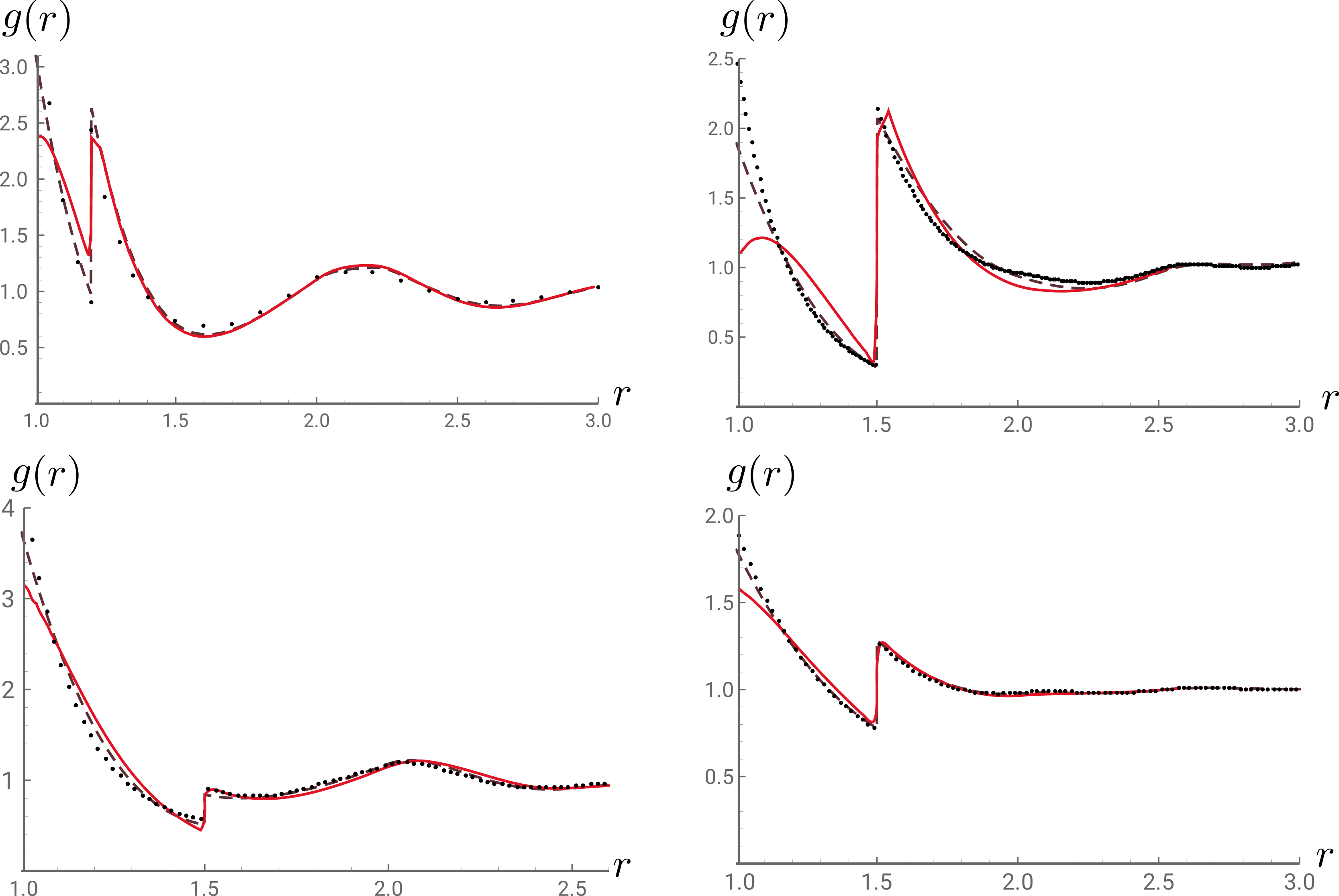}
			\end{center}
			\caption{Pair correlation functions.
			The dashed line is the Rational Fraction Approximation ansatz from \cite{Yuste11,Haro16},
			dots are results from various simulations \cite{Lang99,Zhou09,Guillen10}, the solid line is the high-temperature expansion.
			The top-left graph is for $\lambda=1.2$, $T=1$, $\varphi=0.4$; the top-right graph is for $\lambda = 1.5$, $T=0.5$, $\varphi=0.2094$;
			the bottom-left graph is for $\lambda = 1.5$, $T=2$, $\varphi = 0.4189$; the bottom-right graph is for $\lambda=1.5$, $T=2$, $\varphi = 0.2094$.}
			\label{figgtable}
		\end{figure}

		\end{widetext}

		Finally, we compare our results to those obtained by the Rational Fraction Approximation by Yuste et al. \cite{Yuste11,Haro16}, which are the closest to a
		fully analytical solution without expansions.
		In their work however, they represented the pair correlation function --- which was shown to compare well to numerical data from various simulations \cite{Lang99,Zhou09,Guillen10} ---
		which in our setup requires an additional Fourier transform.
		This causes some numerical artifacts in the vicinity of the discontinuities of $g$ where very large wave number data is required.
		The comparison is shown in Fig.~\ref{figgtable}.

		Let us first stress that the data presented in \cite{Yuste11,Haro16} is mostly in the range where the temperature is neither small nor large, so that
		our expansions are not expected to be very precise.
		Second, following the same line of argument as the one used in the previous section, we chose not to represent the results for low-temperature expansion since for so big
		values of $\lambda$, the outer-core packing fraction is too large for the hard-sphere $T\rightarrow0$ limit to be properly defined.
		With that in mind, the agreement between the high-temperature Percus-Yevick solution and the Rational Fraction Approximation and simulation data appears surprisingly
		satisfactory.
		This can be used to validate the additional set of approximations we used in order to derive fully analytical expressions in the high-temperature limit.

\section{Conclusion}

	All in all, we have shown how Baxter's solution to hard-sphere's structure factor can be extended to treat the high- and low-temperature sectors of the square-shoulder structure factor.
	In the low-temperature regime, our computation can be generalized to higher orders to improve precision.
	Moreover, the typical exponential dependence of $p$ with respect to $\Gamma$ generally ensures that already at moderate temperature the low-temperature predictions should be quite good.
	However, at high temperatures, the lack of simple approximant of the hard-sphere pair correlation function outside of the core makes the Baxter's solution a not so judicious
	starting point for a temperature expansion.
	This result has two main origins: first, the Percus-Yevick closure is particularly adapted to the hard-sphere potential -- it reduces the problem to the computation of $Q(r)$ in a region
	where $g(r)$ is known exactly -- but this cannot be easily generalized to any other potential.
	Second, the Ornstein-Zernike equation involves a convolution integral which appears to be poorly adapted to the design of perturbative expansions other than the virial one (indeed
	the convolution integral is proportional to the density).
	Finally, comparison with numerical data showed reasonable agreement.

\section*{Acknowledgments}

	This work was funded by DAAD.
	Both authors thank T. Kranz for careful reading of the manuscript.
	We thank A. Santos and N. Gadjisade for the numerical data they provided.

\appendix

\section{Low temperature equations - Exact solution}
\label{ALTex}

	Let us first rewrite the Ornstein-Zernike equation Eq.~(\ref{eqLT2}) in the following form :
	\begin{equation}
	\label{AeqLT2}
		\left\{\begin{split}
			& \Psi_1^{(3)}(r) = A\,\Psi_2(r)+B\,\Psi_2'(r) \\
			& \Psi_2^{(3)}(r) = -A\,\Psi_1(r)-B\,\Psi_1'(r)\ ,
		\end{split}\right.
	\end{equation}
	where $\Psi_1(r)=Q_{III}(r)$, $\Psi_2(r)=Q_I(r+R)$, $A=12p\varphi/R^3$ and $B=12p\varphi/R$.
	Both $\Psi_1$ and $\Psi_2$ are solutions of the equation:
	\begin{equation}
		y^{(6)}+B^2\,y''+2AB\,y'+A^2\,y=0\ ,
	\end{equation}
	whose characteristic polynomial is:
	\begin{equation}
	\label{AeqCharPolLT}
		\begin{split}
			& X^6+B^2X^2+2AB\,X+A^2 \\
			&\qquad\qquad\qquad =\big(X^3+\text{i}(BX+A)\big)\big(X^3-\text{i}(BX+A)\big)\ ,
		\end{split}
	\end{equation}
	its roots are thus known.
	The general solution is a combination of exponentials of the six roots, which can be divided in three sets of complex conjugates $\big\{X_1^\pm,X_2^\pm,X_3^\pm\big\}$,
	with the additional condition that it is a real function (what imposes conditions on the relative coefficients of the exponentials of two conjugate roots).
	Explicitly, these roots are given by:
	\begin{equation}
	\label{AeqXpm}
		\left\{\begin{split}
			& X_1^{\pm}=\frac{1}{3\big(f_{\mp}\big)^{1/3}}\left[\mp\text{i}\frac{12(p\varphi)}{R^2}\,3^{2/3}+3^{1/3}\big(f_{\mp}\big)^{2/3}\right] \\
			& X_2^{\pm}=\frac{\pm\text{i}}{6\big(f_{\mp}\big)^{1/3}}\left[(\pm\sqrt{3}\text{i}+1)\frac{12(p\varphi)}{R^2}\,3^{1/6}\sqrt{3}\right. \\
			&\qquad\left.+3^{1/3}(\pm\text{i}+\sqrt{3})\big(f_{\mp}\big)^{2/3}\right] \\
			& X_3^{\pm}=\frac{1}{6\big(f_{\mp}\big)^{1/3}}\left[(\sqrt{3}\pm\text{i})\frac{12(p\varphi)}{R^2}\,3^{1/6}\sqrt{3}\right. \\
			&\qquad\left.-3^{1/3}(-1\mp\text{i}\sqrt{3})\big(f_{\mp}\big)^{2/3}\right]\ ,
		\end{split}\right.
	\end{equation}
	where
	\begin{equation}
		\begin{split}
			\frac{R^3}{18p}\times f_\pm&=\pm3\text{i}\varphi+\sqrt{\pm\text{i}\varphi^2(\pm9\text{i}+16p\varphi)} \\
			&\underset{p\rightarrow0}{=}p\varphi^2\left[-\frac{8}{3}\pm\text{i}\frac{32}{27}(p\varphi)+\frac{256}{243}(p\varphi)^2\right]+O(p^4)\ .
		\end{split}
	\end{equation}
	The signs $\pm$ refer to the factor of Eq.~(\ref{AeqCharPolLT}) the $X$'s are solution of.

	It must not be forgotten that, although this solution is exact, the equation we used in the beginning Eq.~(\ref{AeqLT2}) is only meaningful in the low-temperature regime where the approximation
	that $g(r)$ is constant in the outer-core is justified; it should not be understood in any way as an exact solution to the Wiener-Hopf function problem in presence of a full square shoulder potential.

	We shall now derive the form of the general solution to Eq.~(\ref{AeqLT2}) at order $O(p)$.
	The low-temperature expansion of Eq.~(\ref{AeqXpm}) yields:
	\begin{equation}
		\left\{\begin{split}
			& X_1^\pm \underset{p\rightarrow0}{=}\pm\text{i}\,\frac{2(p\varphi)^{1/3}}{R}\left(\frac{3}{2}\right)^{1/3} \\
			&\qquad\mp\,\frac{2(p\varphi)^{2/3}}{R}\left(\frac{2}{3}\right)^{1/3}+O(p^{4/3}) \\
			& X_2^\pm \underset{p\rightarrow0}{=}\left(\frac{\sqrt{3}\pm\text{i}}{2}\right)\,\frac{2(p\varphi)^{1/3}}{R}\left(\frac{3}{2}\right)^{1/3} \\
			&\qquad+\left(\frac{1\pm\sqrt{3}\text{i}}{2}\right)\,\frac{2(p\varphi)^{2/3}}{R}\left(\frac{2}{3}\right)^{1/3}+O(p^{4/3}) \\
			& X_3^\pm \underset{p\rightarrow0}{=}\left(\frac{-\sqrt{3}\mp\text{i}}{2}\right)\,\frac{2(p\varphi)^{1/3}}{R}\left(\frac{3}{2}\right)^{1/3} \\
			&\qquad	+\left(\frac{-1\mp\sqrt{3}\text{i}}{2}\right)\,\frac{2(p\varphi)^{2/3}}{R}\left(\frac{2}{3}\right)^{1/3}+O(p^{4/3})\ .
		\end{split}\right.
	\end{equation}
	As can be anticipated from the form of the characteristic polynomial Eq.~(\ref{AeqCharPolLT}), it is expressed in terms of the sixth root of (-1) and (+1).

	We now go back to the Wiener-Hopf function. Let's define $X_4=X_1^-$, $X_5=X_2^-$, $X_6=X_3^-$, and the following coefficients:
	\begin{equation}
		Q_{III}(r)=\sum_{i=1}^6 Y_i^{III}\,e^{X_ir}, \ Q_{I}(r)=\sum_{i=1}^6 Y_i^{I}\,e^{X_ir}\ .
	\end{equation}
	The Ornstein-Zernike equation Eq.~ (\ref{eqLT2}) imposes:
	\begin{equation}
		\left\{\begin{split}
			&+\text{i}Y^I_j-Y^{III}_j=0 \ , \quad j\leqslant3\\
			&-\text{i}Y^I_j-Y^{III}_j=0 \ , \quad j\geqslant4\ ,
		\end{split}\right.
	\end{equation}
	In the following, we will thus only work with the set $\{Y_j^{III}\}$ that we will simply denote $\{Y_j\}$.

	Moreover, the Wiener-Hopf function must be real, therefore,
	\begin{equation}
		Y_i=Y_{i+3}^*\ .
	\end{equation}

	Then, we introduce some more adapted notations:
	\begin{equation}
		\alpha_j=Y_j+Y_j^*\,,\ \gamma_j=\text{i}(Y_j-Y_j^*)\ ,
	\end{equation}
	\begin{equation}
		\xi_1=\frac{2(\varphi)^{1/3}}{R}\left(\frac{3}{2}\right)^{1/3}\,,\ \xi_2=\frac{2(\varphi)^{2/3}}{R}\left(\frac{2}{3}\right)^{1/3}\ ,
	\end{equation}
	as a function of which the general solution of Eq.~(\ref{eqLT2}) can be expanded in powers of $p$:
	\begin{equation}
	\label{AeqQp}
		\left\{\begin{split}
			& Q_I(r) \underset{p\rightarrow0}{=} (\gamma_1+\gamma_2+\gamma_3)-\frac{p^{1/3}}{2}\big(\xi_1(2\alpha_1+\alpha_2+\alpha_3)\\
			&\qquad +\xi_1\sqrt{3}(\gamma_3-\gamma_2) +p^{1/3}\xi_2\sqrt{3}(\alpha_2+\alpha_3)\\
			&\qquad +p^{1/3}\xi_2(2\gamma_1-\gamma_2+\gamma_3)\big)r\\
			&\qquad +\frac{p^{2/3}}{4}\xi_1\big(\xi_1(\gamma_2+\gamma_3-2\gamma_1)+\sqrt{3}\xi_1(\alpha_3-\alpha_2) \\
			&\qquad	+4\xi_2 p^{1/3}(\alpha_1-\alpha_2+\alpha_3)\big)r^2\\
			&\qquad	+(\xi_1)^3\frac{p}{6}(\alpha_1-\alpha_2-\alpha_3)r^3+O(p^{4/3}) \\[0.3cm]
			& Q_{III}(r) \underset{p\rightarrow0}{=} (\alpha_1+\alpha_2+\alpha_3)+\frac{p^{1/3}}{2}\big(\xi_1(2\gamma_1+\gamma_2+\gamma_3)\\
			&\qquad +\xi_1\sqrt{3}(\alpha_2-\alpha_3) +p^{1/3}\xi_2\sqrt{3}(\gamma_2+\gamma_3)\\
			&\qquad +p^{1/3}\xi_2(\alpha_2-\alpha_3-2\alpha_1)\big)r\\
			&\qquad +\frac{p^{2/3}}{4}\xi_1\big(\xi_1(\alpha_2+\alpha_3-2\alpha_1)+\sqrt{3}\xi_1(\gamma_2-\gamma_3)\\
			&\qquad	-4\xi_2 p^{1/3}(\gamma_1-\gamma_2+\gamma_3)\big)r^2\\
			&\qquad	+(\xi_1)^3\frac{p}{6}(\gamma_2+\gamma_3-\gamma_1)r^3+O(p^{4/3})
		\end{split}\right.
	\end{equation}

	All in all, even if the general solution to the Ornstein-Zernike equation within our set of approximations is a sum of exponential functions,
	polynomial expressions for the Wiener-Hopf function are recovered in the frame of the $p$-expansion.
	Interestingly, $Q_I$ and $Q_{III}$, when developed at order $O(p)$ see their degree simply be augmented by 1.

	Finally, the general solution to the equation at first order in $p$ can be expressed as a degree three polynomial, what justifies the ansatz Eq.~(\ref{eqQLT}).
	Moreover, Eq.~(\ref{AeqQp}) specifies the $p$ dependence of such coefficients when retrieved in an expansion of the exponential solution.
	Plugging it back into the Ornstein-Zernike equation Eq.~(\ref{eqLT1}), we see that in the boundary condition equation, only the constant term should be taken
	into account in the integral terms with a $p$ prefactor (the coefficients of higher order terms all vanish in the $p\rightarrow0$ limit and are thus subdominant at this order).
	Only then should we plug back the ansatz Eq.~(\ref{eqQLT}) and express the boundary conditions in powers of $p$.

\section{High temperature -- the full Percus-Yevick Solution}
\label{AHTPYA}

	A first way to tackle the problem is to use the Percus-Yevick equation inside the soft core:
	\begin{equation}
		c(r)=\big(1-e^{\Gamma}\big)g(r)\ ,
	\end{equation}
	combined with the Ornstein-Zernike equation for the direct correlation function:
	\begin{equation}
	\label{eqAOZc}
		rc(r)=-Q'(r)+\frac{12\varphi}{R}\,\int_r^dQ'(s)Q(s-r)ds\ .
	\end{equation}
	Plugging this back into Eq.~ (\ref{eqHT1}) to replace every occurrence of $g(r)$ yields the following three equations:
	\begin{widetext}
	\begin{itemize}
		\item $r\in[R;d]$:
			\begin{equation}
			\label{eqAQ3f}
				\begin{split}
					-r=&+\left(\frac{e^{\Gamma}}{1-e^\Gamma}\right)Q_{III}'(r)+\frac{12\varphi}{R^3}\int_0^dQ(s)(s-r)ds-\frac{12\varphi}{R^3(1-e^\Gamma)}\int_0^{r-R}Q_I(s)Q'_{III}(r-s)ds\\
					&+\left(\frac{12\varphi}{R^3}\right)^2\frac{1}{1-e^\Gamma}\int_0^{r-R}Q_I(s)\int_{r-s}^d du\,Q_{III}'(u)Q_I(u-r+s)ds\ .
				\end{split}
			\end{equation}
			
		\item $r\in[d-R;R]$:
			\begin{equation}
			\label{eqAQ2f}
				-r=-Q_{II}'(r)+\frac{12\varphi}{R^3}\int_0^dQ(s)(s-r)ds\ .
			\end{equation}
			
		\item $r\in[0;d-R]$:
			\begin{equation}
			\label{eqAQ1f}	
				\begin{split}
					-r=&-Q_{I}'(r)+\frac{12\varphi}{R^3}\int_0^dQ(s)(s-r)ds-\frac{12\varphi}{R^3(1-e^\Gamma)}\int_{r+R}^dQ_{III}(s)Q'_{III}(r-s)ds\\
					&+\left(\frac{12\varphi}{R^3}\right)^2\frac{1}{1-e^\Gamma}\int_{r+R}^dQ_{III}(s)\int_{s-r}^d du\,Q_{III}'(u)Q_I(u-r+s)ds\ .
				\end{split}
			\end{equation}
	\end{itemize}
	\end{widetext}

	The equation Eq.~ (\ref{eqAQ2f}) still has the usual solution:
	\begin{equation}
		Q_{II}(r)=a_{II}\frac{r^2}{2}+b_{II}\,r+c_{II}\ .
	\end{equation}
	For the two remaining equations, we must explicitly use the $\Gamma$ expansion.
	In the limit $\Gamma\rightarrow0$, the solutions are:
	\begin{equation}
		Q_{III}(r)\underset{\Gamma\rightarrow0}{=}O(\Gamma)\,,\quad Q_I(r)\underset{\Gamma\rightarrow0}{=}Q_{b}(r)+O(\Gamma)\ .
	\end{equation}
	A main difference between the equations (\ref{eqAQ1f}) and (\ref{eqAQ3f}) on the one hand, and the low-temperature (\ref{eqLT2}) on the other hand,
	is the highly non-linear character of the former.
	As a consequence, in order to get linear differential equations on the Wiener-Hopf function, we must not take two, but five additional derivatives.
	With the following notations:
	\begin{equation}
		\upsilon_0(\Gamma)=\frac{1}{1-e^{-\Gamma}}\geqslant0\; ,\ \upsilon_1(\Gamma)=\frac{1}{e^\Gamma-1}\geqslant0\; ,
	\end{equation}
	\begin{equation}
		\mathcal{A}=\frac{12\varphi}{R^3}>0\;,
	\end{equation}
	\begin{equation}
		\omega_0=-\mathcal{A}\int_0^ds\,Q(s)ds\geqslant0\ ,\quad\omega_1=-\mathcal{A}\int_0^d Q(s)ds\ ,
	\end{equation}
	the equation Eq.~ (\ref{eqAQ3f}) becomes:
	\begin{equation}
	\label{eqAQ3fG}
		\begin{split}
			0=&-\upsilon_0\,Q_{III}^{(6)}+\mathcal{A}\,\upsilon_1\big(c_{b}\,Q_{III}^{(5)} +b_{b}\,Q_{III}^{(4)} +a_{b}\,Q^{(3)}_{III}\big)\\
			&+\mathcal{A}^2\upsilon_1\big[c_{b}^2\,Q_{III}^{(4)}+b_{b}^2\,Q_{III}''+a_{b}^2\,Q_{III} \\
			&+2a_{b}b_{b}\,Q_{III}'+2a_{b}c_{b}\,Q_{III}''+2b_{b}c_{b}\,Q_{III}^{(3)}\big]+O(\Gamma^2)\ .
		\end{split}
	\end{equation}

	It is a linear differential equation, whose characteristic polynomial is:
	\begin{equation}
		-\upsilon_0\,X^6+\upsilon_1\,P_b(X)\,X^3+\upsilon_1\,P_b(X)^2\ ,
	\end{equation}
	where we defined:
	\begin{equation}
		P_b(X)=\mathcal{A}\big(c_{b}\,X^2+b_{b}\,X+a_{b}\big)\ .
	\end{equation}
	Defining $\upsilon=\upsilon_0/\upsilon_1$, its roots can be found.
	Indeed, for arbitrary $\theta$, the roots of
	\begin{equation}
		-\upsilon \mathcal{X}^2+\theta \mathcal{X}+\theta^2
	\end{equation}
	are:
	\begin{equation}
		\mathcal{X}_{\pm}=\frac{\theta}{2\upsilon}\big[1\pm\sqrt{1+4\upsilon}\big]=\theta\,\zeta_\pm\ ,
	\end{equation}
	so that finally,
	\begin{equation}
		\begin{split}
			-\upsilon &X^6+P_b(X)\,X^3+P_b(X)^2 \\
			&=-\upsilon\big(X^3-\zeta_+\,P_b(X)\big)\big(X^3-\zeta_-\,P_b(X)\big)\ .
		\end{split}
	\end{equation}
	The remaining roots of polynomials of order three can then be found exactly.
	They all have the form:
	\begin{equation}
		X_i=X_i^0+\Gamma\,X_i^1+O(\Gamma^2)\ ,
	\end{equation}
	with $X_i^0\neq0$ and $X_i^1\neq0$.
	The $Q_{III}$ function thus expands as:
	\begin{equation}
	\label{eqQQ3Sad}
		Q_{III}(r)=\sum_{i=1}^6Y_ie^{X_i r}=\sum_{i=1}^6Y_ie^{X_i^0 r}\big(1+(X_i^1r)\Gamma\big)+O(\Gamma^2)\ .
	\end{equation}
	Finally, the structure of the $\Gamma$ expansion is not compatible with a polynomial Wiener-Hopf function.
	This is due to the fact that $X_i^0\neq0$, which is in strong contrast to what happened in the low-temperature expansion.

\section{High temperature coefficients}
\label{AHTCo}

\begin{widetext}

	$\bullet$ $Q_{III}$ coefficients

	Let us recall the general form of $Q_{III}$:

	\[ Q_{III}(r)=q_0\,e^{q_1 r}+q_2\,e^{q_3 r}+b_{III} r +c_{III} \ . \]

	The exponents in the exponential terms are:
	\begin{equation}
		q_1 = - \frac{2 (\varphi^2 -2 \varphi +10)}{R (2 \varphi^2 -7 \varphi +5)}\ ,\quad
		q_3 = \frac{6 \varphi }{R(1-\varphi)} \ .
	\end{equation}

	The coefficients $q_0$ and $q_2$ vanish, as expected, in the infinite temperature limit ($\Gamma\rightarrow0$):

	\begin{equation}
		q_0 = \Gamma\, \frac{R^2(5-2\varphi)^2 \varphi}{4 (5 \varphi^3-18 \varphi^2+3 \varphi +10)}
		\,e^{\frac{2(\varphi^2 - 2\varphi +10)}{2\varphi^2 -7 \varphi +5}}\ ,
	\end{equation}
	\begin{equation}
		\begin{split}
		q_2 = \Gamma &\,\frac{R^2 \left( (\varphi-1)^2 (5\varphi^2 -13 \varphi -10)\big((6 \lambda -1) \varphi +1\big)
		e^{\frac{2 \lambda (\varphi^2 - 2 \varphi +10)+4 \varphi}{2 \varphi^2 -7\varphi +5}}
		-9 (5-2 \varphi)^2 \varphi^3 e^{\frac{2 (\varphi^2 + 10)}{2 \varphi^2 - 7\varphi +5}}\right)}
		{36 \varphi^2 (5 \varphi^3 - 18 \varphi^2 +3 \varphi +10)} \\
		& \times\,\exp \left(-\frac{\lambda (-10 \varphi^2 +26 \varphi +20)+4 \varphi}{2 \varphi^2
		-7 \varphi + 5}\right) \ .
		\end{split}
	\end{equation}

	Finally, the polynomial part of $Q_{III}$ is parametrized by:
	\begin{equation}
		b_{III} = \Gamma \,\frac{R (1 - \varphi)}{6 \varphi }\ , \quad
		c_{III} = \Gamma \, \frac{R^2 (1 - \varphi)^2}{36 \varphi^2}\ ,
	\end{equation}
	and also vanish at high temperatures.

	$\bullet$ $Q_{II}$ coefficients \\

	They write as $a_{II} = a_b + a_{II}^{(1)}\,\Gamma$, $b_{II} = b_b + b_{II}^{(1)}\,\Gamma$,
	and $c_{II} = c_b + c_{II}^{(1)}\,\Gamma$, where $a_b$, $b_b$ and $c_b$ are the Baxter's values for the corresponding
	hard sphere system given by Eq.~(\ref{eqCOBaxter}).
	The first order coefficients are given below.

	\begin{equation}
		\begin{split}
			a_{II}^{(1)} =& \frac{e^{-\frac{4 (\varphi (3 \varphi +2)+\lambda (\varphi^2 -2 \varphi +10))}{2 \varphi^2
			-7 \varphi +5}}}{3240 (1-\varphi)^3 \varphi^4 (7 \varphi -10) (\varphi^2 -2 \varphi +10)^4
			(5 \varphi^2 -13 \varphi -10)} \\
			&\times \bigg(3645\,e^{\frac{8 (2 \varphi^2 +5)}{2 \varphi^2 -7 \varphi +5}}
			(5-2 \varphi)^6 (40 \varphi^5 -224 \varphi^4 +397 \varphi^3 -139 \varphi^2 -460 \varphi -100) \varphi^8 \\
			& -\!90\, e^{\frac{2 (\varphi^2 +17 \varphi +10 +\lambda (7 \varphi^2 -17 \varphi +10))}
			{(\varphi -1) (2 \varphi -5)}} (5\!-\!2 \varphi)^2 (\varphi^2\! -\!2 \varphi\! +\!10)^4
			(13 \varphi^6\! -1404 \varphi^5\! +3693 \varphi^4\! -2576 \varphi^3\! -153 \varphi^2\! +204\varphi\! -20)
			\varphi^3 \\
			& +90\, e^{\frac{2 (7 \varphi^2 +2 \varphi +10 +\lambda (\varphi^2 -2 \varphi +10))}{(\varphi-1) (2 \varphi-5)}}
			(5-2\varphi )^2 (5 \varphi^2 -13 \varphi -10) \big((378 \lambda^4 +1260 \lambda^3 -3150 \lambda^2
			+6726 \lambda -20245) \varphi^{12} \\
			&\ -3(1062 \lambda^4 +4464 \lambda^3 -7170 \lambda^2 +20912 \lambda -80567)\varphi^{11}\!\!
			+6 (3654 \lambda^4\! +13428 \lambda^3\! -26919 \lambda^2\! +72236 \lambda\! -222189)\varphi^{10} \\
			&\ -2 (45090 \lambda^4 +168318 \lambda^3 -376596 \lambda^2 +928980 \lambda -2125733)\varphi^9 \\
			&\ +9 (33216 \lambda^4 +106008 \lambda^3 -295206 \lambda^2 +643148 \lambda -964925) \varphi^8 \\
			&\ -18 (37860 \lambda^4 +111984 \lambda^3 -360063 \lambda^2 +671954 \lambda -641292)\varphi^7 \\
			&\ +3 (385200 \lambda^4 +938880 \lambda^3 -3373668 \lambda^2 +5337048 \lambda -3319589) \varphi^6 \\
			&\ -36 (34500 \lambda^4 +72300 \lambda^3 -206655 \lambda^2 +226202 \lambda -102809)\varphi^5 \\
			&\ +72 (7500 \lambda^4 +25500 \lambda^3 -2925 \lambda^2 -44980 \lambda +10191) \varphi^4
			-160 (4500 \lambda^3 +9675 \lambda^2 -9870 \lambda -4231) \varphi^3 \\
			&\ +1200 (300 \lambda^2 +530 \lambda -281) \varphi^2 -6000 (20 \lambda +21) \varphi +20000\big) \varphi^3 \\
			& +10\, e^{\frac{38 \varphi +2 \lambda (8 \varphi^2 -19 \varphi +20)}{2 \varphi^2 -7 \varphi +5}}
			(1-\varphi)^2 \big((6 \lambda -1) \varphi +1\big) (\varphi^2 -2 \varphi +10)^4 \\
			&\ \times(65 \varphi^8 -7189 \varphi^7 +36587 \varphi^6 -46849 \varphi^5 -4207 \varphi^4 +28769 \varphi^3
			-1222 \varphi^2 -1780 \varphi+200)\\
			& +e^{\frac{4 (\varphi (3 \varphi +2)+ \lambda (\varphi^2 -2 \varphi +10))}{2 \varphi^2 -7 \varphi +5}}
			(7 \varphi^2\! -17 \varphi\! +10) \big(10 (1620 \lambda^6\! -1296 \lambda^5\! -44550 \lambda^4\!
			+200520 \lambda^3\!-346770 \lambda^2\! +443705)\varphi^{17} \\
			&\ -2 (93960 \lambda^6 -62208 \lambda^5 -2871450 \lambda^4 +13671360 \lambda^3 -25718310 \lambda^2
			+36425363) \varphi^{16} \\
			&\ +4 (378270 \lambda^6 -227448 \lambda^5 -11209590 \lambda^4 +52126380 \lambda^3 -97114815 \lambda^2
			+134693303) \varphi^{15} \\
			&\ -4 (2046060 \lambda^6 -1031616 \lambda^5 -59803110 \lambda^4 +267733800 \lambda^3 -475849665 \lambda^2
			+584842171) \varphi^{14} \\
			&\ +4 (8495280 \lambda^6 -3522528 \lambda^5 -235477530 \lambda^4 +992313540 \lambda^3 -1633492575 \lambda^2
			+1621901438) \varphi^{13} \\
			&\ -4 (27138240 \lambda^6 -8118144 \lambda^5 -702588330 \lambda^4 +2711876940 \lambda^3 -3992650245 \lambda^2
			+2909867434) \varphi^{12} \\
			&\ +10 (26415072 \lambda^6 -3763584 \lambda^5 -623530224 \lambda^4 +2134579824 \lambda^3 -2691937350 \lambda^2
			+1285613177) \varphi^{11} \\
			&\ -4 (123444000 \lambda^6 +6905088 \lambda^5 -2495838420 \lambda^4 +7044079680 \lambda^3
			-7024662225 \lambda^2 +1844967967) \varphi^{10} \\
			&\ +5 (120528000 \lambda^6 +61585920 \lambda^5 -1961348256 \lambda^4 +3642478848 \lambda^3
			-2081782296 \lambda^2 +247814725) \varphi^9 \\
			&\ -10 (45360000 \lambda^6 +60134400 \lambda^5 -314960400 \lambda^4 -618929568 \lambda^3
			+1194566364 \lambda^2 +184924679) \varphi^8 \\
			&\ -80 (2025000 \lambda^6 -10692000 \lambda^5 -75168000 \lambda^4 +268399800 \lambda^3
			-185576751 \lambda^2 -66979673) \varphi^7 \\
			& +400 (810000 \lambda^6 -8626500 \lambda^4 +1634400 \lambda^3 +11216340 \lambda^2 -11624801) \varphi^6 \\
			&\ -320 (1620000 \lambda^5 +1012500 \lambda^4 -12442500 \lambda^3 +2968875 \lambda^2 +531958) \varphi^5 \\
			&\ +8000 (40500 \lambda^4 +36000 \lambda^3 -155925 \lambda^2 -55939) \varphi^4
			-20000 (7200 \lambda^3+5400 \lambda^2 +3991) \varphi^3\\
			&\ +100000 (360 \lambda^2 +553) \varphi^2 +10000000 \varphi -2000000\big)\bigg)\ ,
   		\end{split}
	\end{equation}
	\begin{equation}
		\begin{split}
			b_{II}^{(1)} =& \frac{R\, e^{-\frac{4 (3 \varphi(\varphi+1)+\lambda(\varphi^2-2 \varphi +10))}
			{2 \varphi^2 -7 \varphi +5}}}{19440 (\varphi -1)^3 \varphi^5 (7 \varphi -10) (\varphi^2 -2 \varphi +10)^4
			(5 \varphi^2 -13 \varphi -10)} \\
			&\times \left(10935(5-2 \varphi)^6 (55 \varphi^5 -323 \varphi^4 +643 \varphi^3 -406 \varphi^2 -505 \varphi
			+50) \varphi^9\,e^{\frac{4(4 \varphi^2 +\varphi +10)}{2 \varphi^2-7 \varphi +5}} \right. \\
			& -90(5\!-\!2\varphi)^2 (\varphi^2\!-2 \varphi\! +10)^4 
			e^{\frac{2 (\varphi^2 +19 \varphi +10 +\lambda (7 \varphi^2 -17\varphi +10))}{(\varphi -1)(2\varphi-5)}} \\
			&\times(65 \varphi^7\!\!-4817 \varphi^6 \!+13290 \varphi^5
			\!\!-10843 \varphi^4\! +1108 \varphi^3\! +615 \varphi^2\! -157 \varphi\! +10) \varphi^3  \\
			&+90 e^{\frac{2 (7 \varphi^2 +4 \varphi +10 +\lambda (\varphi^2 -2 \varphi +10))}{2 \varphi^2-7 \varphi +5}}
			(5-2\varphi)^2 (5 \varphi^2-13 \varphi -10) \varphi^3 \\
			&\times \big((1512 \lambda^4 +4788 \lambda^3 -13482 \lambda^2 +26070 \lambda -84971) \varphi^{13}\\
			&\ -2(6561 \lambda^4 +26100 \lambda^3 -49824 \lambda^2 +121383 \lambda-521728) \varphi^{12} \\
			&\ +6 (15147 \lambda^4 +52758 \lambda^3 -123279 \lambda^2 +284553 \lambda -991406) \varphi^{11}\\
			&\ +(-382644 \lambda^4 -1335636 \lambda^3 +3511782 \lambda^2 -7558296 \lambda +19824179) \varphi^{10}\\ 
			&\ +2(642978 \lambda^4 +1909386 \lambda^3 -6261489 \lambda^2 +12232674 \lambda -21465637) \varphi^9 \\
			&\ -9(336096 \lambda^4 +900432 \lambda^3 -3495288 \lambda^2 +6027012 \lambda -6914119) \varphi^8 \\
			&\ +6(883980 \lambda^4 +1900656 \lambda^3 -8714673 \lambda^2 +13149672 \lambda -10213439) \varphi^7 \\
			&\ -3(2041200 \lambda^4 +3468240 \lambda^3 -15629724 \lambda^2 +18185940 \lambda -11287201) \varphi^6 \\
			&\ +36(94500 \lambda^4 +198900 \lambda^3 -405165 \lambda^2 +17586 \lambda -151624) \varphi^5 \\
			&\ +(-540000 \lambda^4 -3168000 \lambda^3 -2329200 \lambda^2 +7987080 \lambda +2474414) \varphi^4 \\
			&\ +20(18000 \lambda^3+88200 \lambda^2 +14220 \lambda -77689) \varphi^3 \\
			&\ -600 (300 \lambda^2 +1080 \lambda +259) \varphi^2 +2000 (30 \lambda +59)\varphi -10000\big)\\
			& +10 e^{\frac{42 \varphi +2 \lambda (8 \varphi^2-19 \varphi +20)}{2 \varphi^2-7 \varphi +5}}
			(1-\varphi)^2 \big((6 \lambda -1) \varphi +1\big)(\varphi^2 -2 \varphi+10)^4\\
			&\times(325 \varphi^9 -24930 \varphi^8 +128421 \varphi^7 -178815 \varphi^6 +13599 \varphi^5 +97101 \varphi^4
			-19860 \varphi^3 -4059 \varphi^2 +1440 \varphi -100)\\
			&+e^{\frac{4 (3 \varphi(\varphi +1)+\lambda (\varphi^2-2 \varphi +10))}{2 \varphi^2-7 \varphi +5}}
			(7 \varphi^2 -17 \varphi +10) \\
			&\times\big(50 (1296 \lambda^6 -1296 \lambda^5 -35802 \lambda^4 +164232 \lambda^3 -285534 \lambda^2
			+387005) \varphi^{18}\\
			&\ -120 (6399 \lambda^6 -5724 \lambda^5-196128\lambda^4+951858\lambda^3-1797306\lambda^2+2718097)\varphi^{17}\\
			&\ +60(104004 \lambda^6 -88128\lambda^5 -3102651 \lambda^4 +14761728 \lambda^3 -27668583 \lambda^2
			+41544184) \varphi^{16}\\
			&\ -18(1902780 \lambda^6 -1474128 \lambda^5 -56018880 \lambda^4 +257529520 \lambda^3 -461962130\lambda^2
			+627546223) \varphi^{15} \\
			&\ +12 (12009060 \lambda^6 -8498088 \lambda^5 -336540285 \lambda^4 +1464498360 \lambda^3 -2445166200 \lambda^2
			+2770845353) \varphi^{14} \\
			&\ -12(39016080\lambda^6 -24311664 \lambda^5 -1025133705 \lambda^4 +4121735040 \lambda^3 -6207195960 \lambda^2
			+5444576974) \varphi^{13} \\
			&\ +3(388385280 \lambda^6 -198402048 \lambda^5 -9367176960 \lambda^4 +33886716960 \lambda^3
			-44403756900 \lambda^2 +28162466203) \varphi^{12} \\
			&\ -12(186604560 \lambda^6 -69304896 \lambda^5 -3915970920 \lambda^4 +12058641720 \lambda^3
			-12981156135 \lambda^2 +5646777676) \varphi^{11} \\
			&\ +3(968112000 \lambda^6 -74711808 \lambda^5 -16900935120 \lambda^4 +38689626240 \lambda^3 
			-29617889220 \lambda^2 +10006489853) \varphi^{10} \\
			&\ -5(483408000 \lambda^6 +198236160 \lambda^5 -4881731328 \lambda^4 +1378324800 \lambda^3
			+5145650136 \lambda^2 +2052733129) \varphi^9 \\
			&\ -120(1620000 \lambda^6 -24537600 \lambda^5 -159615900 \lambda^4 +735475032 \lambda^3
			-614923167 \lambda^2 -107477161) \varphi^8 \\
			& +60(24300000 \lambda^6 -16848000 \lambda^5 -320652000 \lambda^4 +488347200 \lambda^3 
			-96413184 \lambda^2 -224247415) \varphi^7 \\
			&\ -360 (900000\lambda^6 +3960000 \lambda^5 -7380000 \lambda^4 -28832000 \lambda^3 +13139800 \lambda^2
			+340719) \varphi^6 \\
			&\ +480(540000 \lambda^5 +2193750 \lambda^4 -4185000 \lambda^3 -6452250 \lambda^2 -1956989)\varphi^5 \\
			&\ -12000(13500 \lambda^4 +45000 \lambda^3 -30600 \lambda^2 +8477)\varphi^4
			+120000(600 \lambda^3 +1275 \lambda^2 +1703) \varphi^3 \\
			&\ -1200000(15 \lambda^2+2) \varphi^2 -10500000 \varphi +1000000\big)\bigg)\ ,
		\end{split}
	\end{equation}
	\begin{equation}
		\begin{split}
			c_{II}^{(1)} = &-\frac{R^2}{19440 (1 - \varphi)^2 \varphi^5 (7 \varphi -10) (\varphi^2 -2 \varphi +10)^4
			(5\varphi^2 - 13 \varphi -10)}\; e^{-\frac{4(3 \varphi (\varphi +1)+\lambda (\varphi^2 - 2 \varphi +10))}
			{2 \varphi^2 -7 \varphi +5}} \\
			& \times \left( 32805\,(5-2\varphi)^6 (5 \varphi^4 -28 \varphi^3 +54 \varphi^2 -35 \varphi -50) \varphi^9
			e^{\frac{4 (4 \varphi^2 + \varphi +10)}{2 \varphi^2-7 \varphi +5}} \right. \\
			& -90 (5-2 \varphi)^2 (\varphi^2 -2 \varphi +10)^4 (26 \varphi^6 -957 \varphi^5 +2550 \varphi^4
			-2023 \varphi^3 +84 \varphi^2 +87 \varphi -10) \varphi^3
			e^{\frac{2 (\varphi^2 +19 \varphi +10 +\lambda (7 \varphi^2 -17 \varphi +10))}{(\varphi -1)(2 \varphi -5)}}\\
			& +90 e^{\frac{2 (7 \varphi^2 +4 \varphi +10 +\lambda(\varphi^2 -2 \varphi +10))}{(\varphi -1)(2\varphi -5)}}
			(5-2 \varphi)^2 (5 \varphi^2-13 \varphi -10)\varphi^3 
			\big(2 (189 \lambda^4 +504 \lambda^3 -2016 \lambda^2 +2946 \lambda-12118) \varphi^{12} \\
			& \ -3(1062 \lambda^4 +3672 \lambda^3\! -10362 \lambda^2\! +16222 \lambda\! -98039) \varphi^{11}\!\!
			+3 (7308 \lambda^4 +21276 \lambda^3\! -74682 \lambda^2\! +119468 \lambda\! -551639) \varphi^{10} \\
			& \ -4 (22545 \lambda^4 +65475 \lambda^3 -257040 \lambda^2 +406503 \lambda -1353716)\varphi^9 \\
			&\ +9 (33216 \lambda^4 +77184 \lambda^3\! -391584\lambda^2\!+608260 \lambda\! -1273715) \varphi^8 \\
			&\ -18(37860\lambda^4\!+75672 \lambda^3\! -471663\lambda^2\! +693514\lambda\! -896326) \varphi^7 \\
			&\ +3(385200 \lambda^4 +530640 \lambda^3\! -4478364 \lambda^2\! +6127116 \lambda\! -5090155) \varphi^6 \\
			&\ -18(69000 \lambda^4 +55800 \lambda^3 -618630 \lambda^2 +652592 \lambda -415987) \varphi^5 \\
			&\ +18(30000 \lambda^4 +36000 \lambda^3 -156600 \lambda^2 -77660 \lambda -9553) \varphi^4 \\
			&\ -20(18000 \lambda^3 +25200 \lambda^2 -92580 \lambda -13579) \varphi^3
			+600 (300 \lambda^2+380 \lambda -451) \varphi^2 -12000 (5\lambda +4) \varphi +10000\big) \\
			& +10 e^{\frac{42 \varphi +2 \lambda (8 \varphi^2 -19 \varphi +20)}{2 \varphi^2 -7 \varphi +5}} (1-\varphi)^2
			\big((6 \lambda -1) \varphi +1\big) (\varphi^2 -2 \varphi +10)^4 \\
			&\ \times(130 \varphi^8 -5123 \varphi^7 +24931 \varphi^6 -33695 \varphi^5 +1219 \varphi^4 +19573 \varphi^3
			-2021 \varphi^2 -740 \varphi+100) \\
			& +e^{\frac{4 (3 \varphi (\varphi +1)+\lambda (\varphi^2 -2 \varphi +10))}{2 \varphi^2-7 \varphi +5}}
			(7 \varphi^2 -17 \varphi +10) \big(20 (810 \lambda^6 -1296 \lambda^5 -22680 \lambda^4 +109800 \lambda^3
			-193680 \lambda^2 +301955) \varphi^{17}\\
			&\ -2 (93960 \lambda^6 -143856 \lambda^5 -2926530 \lambda^4 +14999400 \lambda^3 -28746630 \lambda^2
			+50793151)\varphi^{16} \\
			&\ +2 (756540 \lambda^6 -1135296 \lambda^5 -22895460 \lambda^4 +115094160 \lambda^3 -218621970 \lambda^2
			+387418181) \varphi^{15} \\
			&\ -20 (409212 \lambda^6 -594216 \lambda^5 -12245580 \lambda^4 +59626872 \lambda^3 -108393921 \lambda^2
			+175186735) \varphi^{14} \\
			&\ +20(1699056 \lambda^6 -2391120 \lambda^5 -48392073 \lambda^4 +223684020 \lambda^3 -378610254 \lambda^2
			+514451747)\varphi^{13} \\
			&\ -20(5427648 \lambda^6 -7324992 \lambda^5 -145135152 \lambda^4 +622230840 \lambda^3 -950117175 \lambda^2
			+1007708074) \varphi^{12} \\
			&\ +(264150720 \lambda^6 -335798784 \lambda^5 -6492921120 \lambda^4 +25178139360 \lambda^3
			-33450806700 \lambda^2 +25867546039) \varphi^{11}\\
			&\ +(-493776000 \lambda^6 +578721024 \lambda^5 +10548668880 \lambda^4 -34996605120 \lambda^3
			+38027120220 \lambda^2 -20070008909) \varphi^{10}\\
			&\ +5 (120528000 \lambda^6 -113840640 \lambda^5 -2146782528 \lambda^4 +5287018176 \lambda^3 
			-3919962600 \lambda^2 +1401192947) \varphi^9 \\
			&\ -20 (22680000 \lambda^6 -12182400 \lambda^5 -211296600 \lambda^4 -48778992 \lambda^3 +474553638 \lambda^2
			-39460333) \varphi^8 \\
			&\ -20 (8100000 \lambda^6 -31104000 \lambda^5 -266976000 \lambda^4 +1143273600 \lambda^3 -988064352 \lambda^2
			+8993921) \varphi^7 \\
			&\ +40(8100000 \lambda^6 -9720000 \lambda^5 -88695000 \lambda^4 +111852000 \lambda^3 +12922200 \lambda^2
			-51988453) \varphi^6\\
			&\ -160 (1620000 \lambda^5 -506250 \lambda^4 -18180000 \lambda^3 +8520750 \lambda^2 +561733) \varphi^5 \\
			&\ +8000(20250 \lambda^4 +4500 \lambda^3 -89775 \lambda^2 -41162) \varphi^4
			-40000 (1800 \lambda^3+675 \lambda^2 +1414) \varphi^3 \\
			&\ +100000 (180 \lambda^2 +359) \varphi^2 +3500000 \varphi -1000000)\bigg)\ .
		\end{split}
	\end{equation}

	$\bullet$ $Q_{I}$ coefficients \\

	$Q_I$ contains both a polynomial and an exponential parts.
	We recall its form:
	\[ Q_I(r)=g_I\frac{r^4}{24}+e_I\frac{r^3}{6}+a_I \frac{r^2}{2}+b_I\, r+c_I+ q_{10}\,e^{q_1r}
	+ q_{11}\,e^{-q_1r} +q_{30}\,e ^{q_3r}\ .\]
	The coefficients $q_1$ and $q_3$ are inherited from $Q_{III}$ and given above.
	Only the three coefficients present in the hard sphere solutions have a non-vanishing
	$\Gamma\rightarrow0$ limit.
	They write as $a_{I} = a_b + a_{I}^{(1)}\,\Gamma$, $b_{I} = b_b + b_{I}^{(1)}\,\Gamma$,
	and $c_{I} = c_b + c_{I}^{(1)}\,\Gamma$, where $a_b$, $b_b$ and $c_b$ are the Baxter's values for the corresponding
	hard sphere system given by Eq.~(\ref{eqCOBaxter}).
	The values of the first order corrections are as follows:

	\begin{equation}
		\begin{split}
			a_{I}^{(1)} =& \frac{e^{-\frac{4 (\varphi (3 \varphi +2)+\lambda (\varphi^2-2 \varphi +10))}{2 \varphi^2 -7 \varphi +5}}}{3240 (1 -\varphi)^3 \varphi^4 (7 \varphi -10) (\varphi^2 -2 \varphi +10)^4
			(5 \varphi^2 -13 \varphi -10)} \\
			&\times \bigg(3645\,e^{\frac{8 (2 \varphi^2+5)}{2 \varphi^2 -7 \varphi +5}} (5-2 \varphi)^6 (40 \varphi^5 -224 \varphi^4 +397 \varphi^3 -139 \varphi^2 -460 \varphi -100) \varphi^8 \\
			&-90\,e^{\frac{2 (\varphi^2+17 \varphi +10 +\lambda (7 \varphi^2 -17 \varphi +10))}{(\varphi-1) (2\varphi-5)}}(5\!-2 \varphi)^2 (\varphi^2\!-2 \varphi\! +10)^4
			(13 \varphi^6\! -1404 \varphi^5\! +3693 \varphi^4\! -2576 \varphi^3\! -153 \varphi^2\! +204 \varphi\! -20) \varphi^3 \\
			&+90\,e^{\frac{2 (7 \varphi^2 +2 \varphi +10 +\lambda (\varphi^2 -2 \varphi +10))}{(\varphi -1) (2 \varphi -5)}}
			(5-2 \varphi)^2(5 \varphi^2 -13 \varphi -10) \big((378 \lambda^4 +1260 \lambda^3 -3150 \lambda^2 +6726 \lambda -20119) \varphi^{12} \\
			&\ -3 (1062 \lambda^4 +4464 \lambda^3 -7170 \lambda^2 +20912 \lambda -80129)\varphi^{11} 
			+6 (3654 \lambda^4 +13428 \lambda^3 -26919 \lambda^2 +72236 \lambda -220596) \varphi^{10} \\
			&\ -4 (22545 \lambda^4 +84159 \lambda^3 -188298 \lambda^2 +464490 \lambda -1051432)\varphi^9 \\ 
			&\ +9 (33216 \lambda^4 +106008 \lambda^3 -295206 \lambda^2 +643148 \lambda -946361) \varphi^8 \\
			&\ -18 (37860 \lambda^4 +111984 \lambda^3 -360063 \lambda^2 +671954 \lambda -615930)\varphi^7 \\
			&\ +3 (385200 \lambda^4 +938880 \lambda^3 -3373668 \lambda^2 +5337048 \lambda -3006533) \varphi^6 \\
			&\ -36 (34500 \lambda^4 +72300 \lambda^3 -206655 \lambda^2 +226202 \lambda -63599)\varphi^5 \\
			&\ +72 (7500 \lambda^4 +25500 \lambda^3 -2925 \lambda^2 -44980 \lambda +29541) \varphi^4 \\
			&\ -80 (9000 \lambda^3 +19350 \lambda^2 -19740 \lambda +1213) \varphi^3 +1200 (300 \lambda^2 +530 \lambda -131) \varphi^2 -6000 (20 \lambda +21) \varphi +20000 \big) \varphi^3 \\
			&+10\, e^{\frac{38 \varphi +2 \lambda (8 \varphi^2 -19 \varphi +20)}{2 \varphi^2 -7 \varphi +5}} (1 -\varphi)^2 \big((6 \lambda -1) \varphi +1\big) (\varphi^2 -2 \varphi +10)^4 \\
			&\times(65 \varphi^8 -7189 \varphi^7 +36587 \varphi^6 -46849 \varphi^5 -4207 \varphi^4 +28769 \varphi^3 -1222 \varphi^2 -1780 \varphi+200) \\
			&+e^{\frac{4 (\varphi (3 \varphi +2)+\lambda (\varphi^2 -2 \varphi +10))}{2 \varphi^2 -7 \varphi +5}} (7 \varphi^2 -17 \varphi +10)
			\big(10 (1620 \lambda^6 -1296 \lambda^5 -44550 \lambda^4 +200520 \lambda^3 -345150 \lambda^2 +448475) \varphi^{17} \\
			&\ -2 (93960 \lambda^6 -62208 \lambda^5 -2871450 \lambda^4 +13671360 \lambda^3 -25608150 \lambda^2 +36781223) \varphi^{16} \\
			&\ +4 (378270 \lambda^6 -227448 \lambda^5 -11209590 \lambda^4 +52126380 \lambda^3 -96638535 \lambda^2 +136214483) \varphi^{15} \\
			&\ -4 (2046060 \lambda^6 -1031616 \lambda^5 -59803110 \lambda^4 +267733800 \lambda^3 -473000085 \lambda^2 +593797486) \varphi^{14} \\
			&\ +4 (8495280 \lambda^6 -3522528 \lambda^5 -235477530 \lambda^4 +992313540 \lambda^3 -1620526905 \lambda^2 +1661028218) \varphi^{13} \\
			&\ -4 (27138240 \lambda^6 -8118144 \lambda^5 -702588330 \lambda^4 +2711876940 \lambda^3 -3946475385 \lambda^2 +3041213389) \varphi^{12} \\
			&\ +10 (26415072 \lambda^6 -3763584 \lambda^5 -623530224 \lambda^4 +2134579824 \lambda^3 -2640413574 \lambda^2 +1421349431) \varphi^{11} \\
			&\ -4 (123444000 \lambda^6 +6905088 \lambda^5 -2495838420 \lambda^4 +7044079680 \lambda^3 -6742004625 \lambda^2 +2510757187) \varphi^{10} \\
			&\ +5 (120528000 \lambda^6 +61585920 \lambda^5 -1961348256 \lambda^4 +3642478848 \lambda^3 -1710913752 \lambda^2 +999963877) \varphi^9 \\
			&\ -10 (45360000 \lambda^6 +60134400 \lambda^5 -314960400 \lambda^4 -618929568 \lambda^3 +1409831964 \lambda^2 +520834991) \varphi^8 \\
			&\ -80 (2025000 \lambda^6 -10692000 \lambda^5 -75168000 \lambda^4 +268399800 \lambda^3 -202424751 \lambda^2 -78917885) \varphi^7 \\
			&\ +400 (810000 \lambda^6 -8626500 \lambda^4 +1634400 \lambda^3 +11702340 \lambda^2 -8366171) \varphi^6 \\
			&\ -320 (1620000 \lambda^5 +1012500 \lambda^4 -12442500 \lambda^3 +5500125 \lambda^2 +4354708) \varphi^5 \\
			&\ +8000 (40500 \lambda^4 +36000 \lambda^3 -115425 \lambda^2 -30289) \varphi^4 \\
			&\ -20000 (7200 \lambda^3 +5400\lambda^2 -59) \varphi^3 +100000 (360 \lambda^2+ 373) \varphi^2 +10000000 \varphi -2000000 \big)\bigg)\ ,
		\end{split}
	\end{equation}
	\begin{equation}
		\begin{split}
			b_{I}^{(1)} =& \frac{R\,e^{-\frac{4 (3 \varphi (\varphi +1)+\lambda (\varphi^2 -2 \varphi +10))}{2 \varphi^2
			-7 \varphi +5}}}{19440 (\varphi -1)^3\varphi^5(7\varphi-10)(\varphi^2-2\varphi+10)^4(5\varphi^2-13\varphi-10)} \\
			&\times \bigg(10935\,e^{\frac{4 (4 \varphi^2 +\varphi +10)}{2 \varphi^2 -7 \varphi +5}}
			(5-2 \varphi)^6 (55 \varphi^5 -323 \varphi^4 +643 \varphi^3 -406 \varphi^2 -505 \varphi +50) \varphi^9\\
			&-90\,e^{\frac{2 (\varphi^2 +19 \varphi +10 +\lambda (7 \varphi^2 -17\varphi +10))}{(\varphi -1)(2 \varphi -5)}}
			(5-2\varphi)^2(\varphi^2-2\varphi+10)^4 \\
			&\times (65 \varphi^7-4817 \varphi^6+13290 \varphi^5-10843 \varphi^4+1108 \varphi^3+615\varphi^2
			-157 \varphi +10) \varphi^3\\
			&+90\,e^{\frac{2 (7 \varphi^2+4 \varphi +10 +\lambda (\varphi^2-2 \varphi +10))}{2 \varphi^2-7 \varphi +5}}
			(5-2 \varphi)^2 (5 \varphi^2-13 \varphi-10) \big((1512 \lambda^4 +4788 \lambda^3 -13482 \lambda^2
			+26826 \lambda -84089) \varphi^{13} \\
			&\ -2 (6561 \lambda^4 +26100 \lambda^3 -49824 \lambda^2 +125325 \lambda -516499)\varphi^{12} \\
			&\ +6 (15147 \lambda^4 +52758 \lambda^3 -123279 \lambda^2 +294111 \lambda -979388) \varphi^{11} \\
			&\ +(-382644 \lambda^4 -1335636 \lambda^3 +3511782 \lambda^2 -7832724 \lambda +19486679) \varphi^{10} \\
			&\ +2 (642978 \lambda^4 +1909386 \lambda^3 -6261489 \lambda^2 +12733902 \lambda -20894452) \varphi^9 \\
			&\ -9 (336096 \lambda^4 +900432 \lambda^3 -3495288 \lambda^2 +6331356\lambda -6594541) \varphi^8 \\
			&\ +6 (883980 \lambda^4 +1900656 \lambda^3 -8714673 \lambda^2 +14088840 \lambda -9326291) \varphi^7 \\
			&\ -3 (2041200 \lambda^4 +3468240 \lambda^3 -15629724 \lambda^2 +21009060 \lambda -8950585) \varphi^6 \\
			&\ +36 (94500 \lambda^4 +198900 \lambda^3 -405165 \lambda^2 +249786 \lambda +23486) \varphi^5 \\
			&\ -2 (270000 \lambda^4 +1584000 \lambda^3 +1164600 \lambda^2 -1671540 \lambda +588893) \varphi^4 \\
			&\ +20 (18000 \lambda^3 +88200 \lambda^2 +68220 \lambda -16489) \varphi^3
			-600 (300 \lambda^2 +1080 \lambda +559) \varphi^2 +2000 (30 \lambda +59)\varphi -10000\big) \varphi^3\\
			&+10\,e^{\frac{42 \varphi +2 \lambda (8 \varphi^2 -19 \varphi +20)}{2 \varphi^2 -7 \varphi +5}}
			(1 -\varphi)^2 \big((6 \lambda -1) \varphi +1\big) (\varphi^2 -2 \varphi+10)^4 \\
			&\times (325 \varphi^9 -24930 \varphi^8 +128421 \varphi^7 -178815 \varphi^6 +13599 \varphi^5
			+97101 \varphi^4 -19860 \varphi^3 -4059 \varphi^2 +1440 \varphi -100)\\
			& +e^{\frac{4 (3 \varphi(\varphi +1)+\lambda (\varphi^2 -2 \varphi +10))}{2 \varphi^2 -7 \varphi +5}}
			(7 \varphi^2 -17 \varphi +10) \\
			&\times \big(50 (1296 \lambda^6 -1296 \lambda^5 -35802 \lambda^4 +165528\lambda^3
			-285210 \lambda^2 +378887)\varphi ^{18} \\
			&\ -60 (12798 \lambda^6 -11448 \lambda^5 -392256 \lambda^4 +1918404 \lambda^3 -3590670 \lambda^2
			+5328215) \varphi^{17} \\
			&\ +60 (104004 \lambda^6 -88128 \lambda^5 -3102651 \lambda^4 +14888736 \lambda^3 -27633159 \lambda^2
			+40626490)\varphi^{16} \\
			&\ -18 (1902780 \lambda^6 -1474128 \lambda^5 -56018880 \lambda^4 +260062480 \lambda^3 -461223050 \lambda^2
			+610036803) \varphi^{15} \\
			&\ +12 (12009060 \lambda^6 -8498088 \lambda^5 -336540285 \lambda^4 +1481785920 \lambda^3
			-2439894450 \lambda^2 +2661335948) \varphi^{14} \\
			&\ -12 (39016080\lambda^6 -24311664 \lambda^5 -1025133705 \lambda^4 +4183301520 \lambda^3
			-6187482450 \lambda^2 +5101090309) \varphi^{13} \\
			&\ +3 (388385280 \lambda^6 -198402048 \lambda^5 -9367176960 \lambda^4 +34573700640 \lambda^3
			-44170444500 \lambda^2 +24913174483) \varphi^{12} \\
			&\ -12 (186604560 \lambda^6 -69304896 \lambda^5 -3915970920 \lambda^4 +12435518520 \lambda^3
			-12844000455 \lambda^2 +4232182486) \varphi^{11} \\
			&\ +3 (968112000 \lambda^6 -74711808 \lambda^5 -16900935120 \lambda^4 +41162083200 \lambda^3
			-28622898180 \lambda^2 +3256488773) \varphi^{10} \\
			&\ -5 (483408000 \lambda^6 +198236160 \lambda^5 -4881731328 \lambda^4 +3100449600 \lambda^3
			+5947049880 \lambda^2 -647233895) \varphi^9 \\
			&\ -120 (1620000 \lambda^6 -24537600 \lambda^5 -159615900 \lambda^4 +690547032 \lambda^3
			-644093967 \lambda^2 -110042995) \varphi^8 \\
			&\ +60 (24300000 \lambda^6 -16848000 \lambda^5 -320652000 \lambda^4 +501307200 \lambda^3
			-115637184 \lambda^2 -349863331) \varphi^7 \\
			&\ -360 (900000 \lambda^6 +3960000 \lambda^5 -7380000 \lambda^4 -19832000 \lambda^3 +15929800 \lambda^2
			-10078081) \varphi^6 \\
			&\ +480 (540000 \lambda^5 +2193750 \lambda^4 -1485000 \lambda^3 -4089750 \lambda^2 -1680989) \varphi^5 \\
			&\ -12000 (13500 \lambda^4 +45000 \lambda^3 -3600 \lambda^2 +18827) \varphi^4 \\
			&\ +120000 (600 \lambda^3 +1275 \lambda^2 +878) \varphi^3
			-1200000 (15 \lambda^2 -13) \varphi^2 -10500000 \varphi +1000000\big)\bigg)\ ,
		\end{split}
	\end{equation}
	\begin{equation}
		\begin{split}
			c_{I}^{(1)} =& -\frac{R^2\,e^{-\frac{4 (3 \varphi (\varphi +1)+\lambda (\varphi^2-2 \varphi +10))}
			{2 \varphi^2-7 \varphi +5}}}{19440 (1-\varphi)^2\varphi^5 (7 \varphi -10)(\varphi^2-2\varphi+10)^4
			(5\varphi^2-13 \varphi -10)} \\
			&\times\bigg(32805\,e^{\frac{4 (4 \varphi^2 +\varphi +10)}{2 \varphi^2-7 \varphi +5}}
			(5-2 \varphi)^6 (5 \varphi^4 -28 \varphi^3+54 \varphi^2-35 \varphi -50) \varphi^9 \\
			&-90\,e^{\frac{2 (\varphi^2+19 \varphi+10 +\lambda (7\varphi^2-17 \varphi +10))}{(\varphi-1)(2 \varphi -5)}}
			(5-2 \varphi)^2 (\varphi^2-2 \varphi +10)^4 (26 \varphi^6\!-957 \varphi^5\!+2550 \varphi^4\!
			-2023 \varphi^3\!+84 \varphi^2\!+87 \varphi\! -10) \varphi^3\\
			&+90\,e^{\frac{2 (7 \varphi^2+4 \varphi +10 +\lambda (\varphi^2-2 \varphi +10))}{(\varphi-1)(2\varphi-5)}}
			(5-2 \varphi)^2(5 \varphi^2-13\varphi -10) \big( (378 \lambda^4+1008 \lambda^3-3654 \lambda^2
			+6774 \lambda -24467) \varphi^{12} \\
			&\ -6(531 \lambda^4+1836 \lambda^3-4587 \lambda^2+9707 \lambda -49197) \varphi^{11}
			+3 (7308 \lambda^4+21276 \lambda^3 -66312 \lambda^2 +140312 \lambda -555391) \varphi^{10} \\
			&\ +(-90180 \lambda^4 -261900 \lambda^3 +916056 \lambda^2 -1900980 \lambda +5486081) \varphi^9 \\
			&\ +18 (16608 \lambda^4 +38592 \lambda^3 -174174 \lambda^2 +352319 \lambda -654915) \varphi^8 \\
			&\ -9 (75720 \lambda^4 +151344 \lambda^3 -834390 \lambda^2 +1610228 \lambda -1903537) \varphi^7 \\
			&\ +12(96300 \lambda^4 +132660 \lambda^3 -966501 \lambda^2 +1807953 \lambda -1438966) \varphi^6 \\
			&\ -18 (69000 \lambda^4 +55800 \lambda^3 -485430 \lambda^2 +857912 \lambda -540831) \varphi ^5 \\
			&\ +72(7500 \lambda^4 +9000 \lambda^3 -14400 \lambda^2 +16810 \lambda -16577) \varphi^4
			-20 (18000 \lambda^3 +52200 \lambda^2 -40380 \lambda -6769) \varphi^3 \\
			&\ +600 (300 \lambda^2 +680 \lambda -111) \varphi^2-6000 (10 \lambda +13) \varphi +10000\big) \varphi^3\\
			&+10\,e^{\frac{42 \varphi +2\lambda (8 \varphi^2-19 \varphi +20)}{2 \varphi^2-7 \varphi +5}}
			(1- \varphi)^2 \big((6 \lambda-1) \varphi +1\big) (\varphi^2 -2 \varphi +10)^4\\
			&\times(130 \varphi^8 -5123 \varphi^7 +24931 \varphi^6 -33695 \varphi^5 +1219 \varphi^4
			+19573 \varphi^3 -2021 \varphi^2 -740 \varphi +100)\\
			&\ +e^{\frac{4 (3 \varphi(\varphi +1)+\lambda (\varphi^2-2 \varphi +10))}{2 \varphi^2 -7 \varphi +5}}
			(7 \varphi^2 -17 \varphi +10) \\
			&\times \big(10 (1620 \lambda^6 -2592 \lambda^5 -42930 \lambda^4 +220680 \lambda^3 -401670 \lambda^2
			+648175) \varphi^{17} \\
			&\ -4 (46980 \lambda^6 -71928 \lambda^5 -1386720 \lambda^4 +7536420 \lambda^3 -14871330 \lambda^2
			+27195773) \varphi^{16} \\
			&\ +2 (756540 \lambda^6 -1135296 \lambda^5 -21619710 \lambda^4 +115729200 \lambda^3 -226753020 \lambda^2
			+416431751) \varphi^{15} \\
			&\ -20 (409212 \lambda^6 -594216 \lambda^5 -11518281 \lambda^4 +60006816 \lambda^3 -112954005 \lambda^2
			+190268218) \varphi^{14} \\
			&\ +20 (1699056 \lambda^6 -2391120 \lambda^5 -45229671 \lambda^4 +225412776 \lambda^3 -397526238 \lambda^2
			+569568254) \varphi^{13} \\
			&\ -10 (10855296 \lambda^6 -14649984 \lambda^5 -268890192 \lambda^4 +1256774976 \lambda^3
			-2020017528 \lambda^2 +2305873787) \varphi^{12} \\
			&\ +(264150720 \lambda^6 -335798784 \lambda^5 -5933865600 \lambda^4 +25521631200 \lambda^3
			-36325062540 \lambda^2 +31313237899) \varphi^{11} \\
			&\ +(-493776000 \lambda^6 +578721024 \lambda^5 +9411778800 \lambda^4 -35750358720 \lambda^3
			+43142335020 \lambda^2 -26932888109) \varphi^{10} \\
			&\ +5 (120528000 \lambda^6 -113840640 \lambda^5 -1817857728 \lambda^4 +5534263872 \lambda^3
			-5153367096 \lambda^2 +2365312889) \varphi^9 \\
			&\ -20 (22680000 \lambda^6 -12182400 \lambda^5 -132078600 \lambda^4 +22976208 \lambda^3 +279039294 \lambda^2
			-32038459) \varphi^8 \\
			&\ -160 (1012500 \lambda^6 -3888000 \lambda^5 -36105750 \lambda^4 +137293200 \lambda^3
			-130040019\lambda^2 +18477964) \varphi^7 \\
			&\ +40 (8100000 \lambda^6 -9720000 \lambda^5 -70470000 \lambda^4 +115092000 \lambda^3 -58708800 \lambda^2
			-15808057) \varphi^6 \\
			&\ -160 (1620000 \lambda^5 +2531250 \lambda^4 -14805000 \lambda^3 +3492000 \lambda^2 -307817)\varphi^5 \\
			& +16000 (10125 \lambda^4 +15750 \lambda^3 -33075 \lambda^2 -19201) \varphi^4
			-20000 (3600 \lambda^3 +4050 \lambda^2 +3863) \varphi^3 \\
			&\ +200000 (90 \lambda^2 +97) \varphi^2 +6500000 \varphi -1000000\big)\bigg)\ .
		\end{split}
	\end{equation}

	The other coefficients are:
	\begin{equation}
		g_{I} = \Gamma\,\frac{2 (\varphi -1)}{R^2} \ ,\quad
		e_I= -\Gamma\, \frac{(\varphi -1)^2}{3 R\, \varphi }\ ,
	\end{equation}
	\begin{equation}
		\begin{split}
		q_{10} =& \Gamma\, \frac{3 R^2 (5-2 \varphi)^4 \varphi^2
		(6 \varphi^4 -29 \varphi^3 +72 \varphi^2 -72 \varphi +50)}
		{16 (\varphi^2-2 \varphi +10)^3 (5 \varphi^3-18 \varphi^2+3 \varphi +10)}\ ,
		\end{split}
	\end{equation}
	\begin{equation}
		\begin{split}
		q_{11} =&\Gamma\, \frac{3 R^2(5-2\varphi)^4\varphi^2}{16 (\varphi-1)^2(7\varphi-10)(\varphi^2-2\varphi+10)^3}
		\times \exp\left(\frac{2 \big(-2\lambda(\varphi^2 -2 \varphi +10)+\varphi^2 -4 \varphi +10\big)}
		{2 \varphi^2 -7 \varphi +5}\right) \\
		&\times \left(2(\varphi-1)^2 \big(2 \lambda(\varphi^2 -2 \varphi+10)+2 \varphi^2 -7 \varphi +5\big)\,
		e^{\frac{2 \lambda (\varphi^2 -2 \varphi +10)+4 \varphi}{2 \varphi^2 -7 \varphi +5}}
		-\varphi(2 \varphi^3 -9 \varphi^2 +30 \varphi -50)\,
		e^{\frac{2(\varphi^2 +10)}{2 \varphi^2 -7 \varphi +5}}\right) \ ,
		\end{split}
	\end{equation}
	\begin{equation}
		\begin{split}
		q_{30} =& \Gamma\,\frac{R^2 (13 \varphi^5 -44 \varphi^4 +163 \varphi^3 -310 \varphi^2 +107 \varphi-10)}
		{648 (\varphi-1)^2 \varphi^4 (7 \varphi -10) (5 \varphi^2 -13 \varphi -10)} 
		\times	\exp\left(\frac{2(\lambda (5 \varphi^2 -13 \varphi -10)+(13-6\varphi) \varphi)}
		{2 \varphi^2 -7 \varphi +5}\right) \\
		&\times	\left((1- \varphi)^2 (5 \varphi^2 -13 \varphi -10) \big((6\lambda -1) \varphi +1\big)
		\,e^{\frac{2 \lambda(\varphi^2 -2 \varphi+10)
		+4\varphi}{2 \varphi^2 -7 \varphi +5}}
		-9\,e^{\frac{2(\varphi^2 +10)}{2 \varphi^2 -7 \varphi +5}} (5-2\varphi)^2\varphi^3\right) \ .
		\end{split}
	\end{equation}

	They all vanish at high temperature, giving back the well known Baxter's form for hard spheres.

\end{widetext}
\bibliography{Struct1.bib}

\end{document}